\documentclass{aa}  

\usepackage{graphicx}
\usepackage{multirow}
\usepackage{amsmath}
\usepackage{subfigure}
\usepackage{rotating}
\usepackage{xcolor}
\usepackage{etoolbox}
\usepackage[flushleft]{threeparttable}

\usepackage{txfonts}
\makeatletter
\def\blfootnote{\xdef\@thefnmark{}\@footnotetext}
\makeatother
\usepackage[colorlinks=true,linkcolor=blue,citecolor=blue]{hyperref}

\usepackage{natbib}

\begin{document}

   \title{The K-band highest-resolution images of the Mira star R Car with GRAVITY-VLTI}

   \author{A. Rosales-Guzman
          \inst{1}
          ,
          J. Sanchez-Bermudez\inst{1,2}
          ,
          C. Paladini\inst{3},
          A. Alberdi\inst{4},
          W. Brandner\inst{2},
          E. Cannon\inst{5},
          G. González-Torà\inst{6,7},
          X. Haubois\inst{3},
          Th. Henning\inst{2},
          P. Kervella\inst{8},
          M. Montarges\inst{8},
          G. Perrin\inst{8},
          R. Schödel\inst{4},
          M. Wittkowski\inst{6}
          }

  \institute{Instituto de Astronomía, Universidad Nacional Autónoma de México, Apdo. Postal 70264, Ciudad de México, 04510, México\\
              \email{jarosales@astro.unam.mx}
         \and
             Max-Planck-Institut für Astronomie, Königstuhl 17, D-69117 Heidelberg, Germany 
         \and
            European Southern Observatory (ESO), Alonso de Córdova 3107, Vitacura, Santiago, Chile 
         \and
             Instituto de Astrofísica de Andalucía, Glorieta de la Astronomía s/n, 18008 Granada, España
        \and 
            Institute of Astronomy, KU Leuven, Celestijnenlaan 200D B2401, 3001 Leuven, Belgium
        \and
            European Southern Observatory (ESO), Karl-Schwarzschild-Str. 2, D-85748 Garching bei München, Germany
        \and 
            Astrophysics Research Institute, Liverpool John Moores University, 146 Brownlow Hill, Liverpool L3 5RF, United Kingdom
        \and 
            LESIA, Observatoire de Paris, Université PSL, CNRS, Sorbonne Université, Université Paris Cité, 5 place Jules Janssen, 92195 Meudon, France
 }

   \date{Received XXXX; accepted YYYY}

 
  \abstract
  {The mass-loss mechanisms in M-type AGB stars are still not well understood, in particular the formation of dust-driven winds from the innermost gaseous layers around these stars. One way to understand the gas-dust interaction in these regions and its impact on the mass-loss mechanisms is through the analysis of high-resolution observations of the stellar surface and its closest environment.}
  {We aim at characterizing the inner circumstellar environment ($\sim$3 R$_*$) of the M-type Mira star R Car in the near-infrared at different phases of a pulsation period.}
  {We used GRAVITY interferometric observations in the $K-$band obtained at two different epochs over 2018. Those data were analyzed using parametric models and image reconstruction of both the pseudo-continuum and the CO band-heads observed. The reported data are the highest angular resolution observations on the source in the K-band.} 
{We determine sizes of R Car's stellar disk of 16.67$\pm 0.05$ mas (3.03 au) in January, 2018 and, 14.84$\pm 0.06$ mas (2.70 au) in February, 2018, respectively. From our physical model, we determined temperatures and size ranges for the innermost CO layer detected around R Car. The derived column density of the CO is in the range $\sim$ 9.18$\times$ 10$^{18}$ - 1$\times$ 10$^{19}$ cm$^{-2}$, sufficient to permit dust nucleation and the formation of stable dust-driven winds. We find that magnesium composites, Mg$_2$SiO$_4$ and MgSiO$_3$, have temperatures and condensation distances consistent with the ones obtained for the CO layer model and pure-line reconstructed images, being them the most plausible dust types responsible of wind formation. Our reconstructed images show evidence of asymmetrical and inhomogeneous structures, which might trace a complex and perhaps clumpy structure of the CO molecule distribution.}  
{Our work demonstrates that the conditions for dust nucleation and thus for initialising dust-driven winds in M-type AGB stars are met in R Car and we identify Magnesium composites as the most probable candidates. We find structural changes between two observing epochs ( which are separated by $\sim$ 10\% of the full pulsation period of the star) and evidence for the effects of asymmetries and clumpiness. These observational evidence is crucial to constrain the role of convection and pulsation in M-type stars. 
}


   \keywords{AGB stars -- individual--R Car --interferometry --
                imaging
               }
\titlerunning{The K-band highest-resolution images of the Mira star R Car with GRAVITY-VLTI} 
\authorrunning{Rosales-Guzmán et al. }

\maketitle
%
\section{Introduction}\label{sec:intro}

When\blfootnote{The calibrated GRAVITY data used in this work are available at the Optical Interferometric Data Base of the JMMC. Examples of the Python code used for optimizing the Uniform Disk model described in Sec.\ref{geo_continuum}, the MOLsphere model in Sec.\ref{sec:moslphere}, the bootstrapping method described in \ref{subsec:reconstruct_imgs} and the PCA analysis described in \ref{subsec:PCA} are available to the reader in the following Github \href{https://github.com/abelrg25/RCar_GRAVITY}{repository}.} low to intermediate mass stars reach the asymptotic giant branch (AGB), they are characterized by an inert C and O nucleus surrounded by He and H burning shells \citep{hofner2008winds}. The early AGB phase begins with He-shell burning and subsequently evolves to a thermal pulse AGB phase where the H-shell burns. At the beginning of this phase, the atmosphere of the star is dominated by O-rich molecules \citep[like CO, TiO, SiO, H$_2$O, VO][]{ wittkowski2014atmosphere} but, eventually (and depending on the mass of the star), the 3rd dredge-up takes place, transporting the C formed in the He burning shell to the atmosphere. Depending on the efficiency of this dredge-up, the star will be of M-type (C/O $\sim$ 0.5), C-type (C/O > 1), or an intermediate star (0.5 < C/O < 1) \citep{hofner2018mass}. 

Models of dust formation in AGBs suggest that pulsations are not sufficient to form the observed outflows and mass-loss rates. However, it is believed that there is a momentum transfer due to propagating shock-waves in the atmosphere that lift gas creating density-enhanced layers around the star where the nucleation of dust can occur \citep{Jeong_2003}. 

The models suggest that the mechanisms for driving dust differ depending on the dust composition. C-type AGBs form amorphous Carbon grains with large opacities that can be accelerated by radiation pressure \citep{gautschy2004dynamic, sacuto2011observing, rau2015modelling, rau2017adventure}. For M-type AGB stars, the high chemical complexity around them makes identifying the dust grains driving the dusty winds more complex. Theoretical models of the formation of dust-driven winds in M-type AGBs suggest that the distance(s) from the photosphere to the innermost molecular layers (located at a few stellar radii) determine how far the shock-waves can levitate the gas. This is strongly coupled with the condensation region where dust of a given chemistry must be formed to produce the observed winds \citep{nowotny2005atmospheric, Bladh_2012, liljegren2016dust}. Therefore, high-angular resolution observations of the innermost molecular layers are particularly important to constrain the gas-dust coupling and the dust chemistry \citep[see e.g.,][]{norris2012close}.

During the last decade, due to the enhanced spectral and spatial resolution made available, near-infrared interferometry has become an important technique to map the innermost structure of AGBs' wind in order to link the observed morphology with the mass-loss models. Most of the analyses based on interferometric data consist of model fitting directly the observables (squared visibilities -V$^2$- and closure phases -CPs-) with parametric models of the star's atmosphere and/or more sophisticated radiative transfer and hydrodynamic simulations with codes like CODEX \citep{ireland2008dynamical, ireland2011dynamical} or CO5BOLD \citep{freytag2002spots, freytag2003co5bold}. Some examples of these studies include \citet[][]{ireland2004multiwavelength}, \citet{Wittkowski_2008, Wittkowski_2011, Wittkowski_2016} and, \citet{haubois2015resolving}.

However, due to the complexity of the environment, it is still difficult to fully reproduce the asymmetries traced by the  CPs of the interferometric data with those models. Therefore, complementary techniques, like aperture-synthesis image reconstruction can be used to better depict the inner circumstellar layers of AGB stars. Examples of interferometric imaging applied to AGB stars include the analysis presented by \citet{wittkowski2017aperture} on the C-rich AGB star R Sculptoris using the $H-$band ($\lambda_0 \sim$ 1.76 $\mu$m) beam-combiner PIONIER \citep{le2011pionier} at the Very Large Telescope Interferometer (VLTI). In that study the reconstructed images are consistent with a dynamic atmosphere and a stellar disk dominated by giant convection cells which has a strong impact on the distribution of the molecular layers and the formation of dust at a few stellar radii. 

A further example includes interferometric imaging applied to the LPV Mira, using the Interferometric Optical Telescope Array (IOTA) in the $H-$band, reported by \citet{perrin2020evidence}. In that study, a large localized absorbing patch is observed in the images. This patch is interpreted as a local dust formation zone and, it is consistent with the combined effect of pulsation and convection in the mass-loss process of AGBs \citep[see also][]{paladini2018large}. These examples highlight the importance of interferometric imaging to confront the observed morphologies with theoretical models of dust-driven winds and mass-loss processes in AGBs. However, there are still limited observational information, in particular for the case of M-type AGBs.

In this work, we aim at characterizing the innermost environment of the M-type Mira star R Car using infrared interferometry, in order to link its morphology with models of the formation of dust-driven winds. For that purpose, we present new $K-$band ($\lambda_0 = 2.2 \mu m$) interferometric data obtained in 2018 with the instrument GRAVITY \citep{abuter2017first} at the VLTI. These data were analyzed to depict the innermost structure of R Car at two phases within one pulsation cycle. Our analyses include: (i) parametric model-fitting of the squared visibilities (V$^2$); (ii) image reconstruction of both the pseudo-continuum and across the first ($\lambda_0 \sim$ 2.29 $\mu$m) and second ( $\lambda_0 \sim$ 2.32 $\mu$m) CO band-heads, and; (iii) an estimation of the CO layer(s) physical parameters such as temperature, size, and optical thickness by using a single-layer model. This paper is structured as follows: observations and data reduction are presented in Section \ref{sec:obs_data}. In Section \ref{analysis}, we describe the results of our geometrical model fitting, our image reconstruction procedure, and our single-layer model. Our results are discussed in Section \ref{sec:discussion}. Finally, we present our conclusions in Section \ref{sec:conclusions}. 

%

\section{Observations and data reduction}\label{sec:obs_data}

\subsection{Properties of R Car}
\begin{figure}[htp]
    \centering
    \includegraphics[width=9cm]{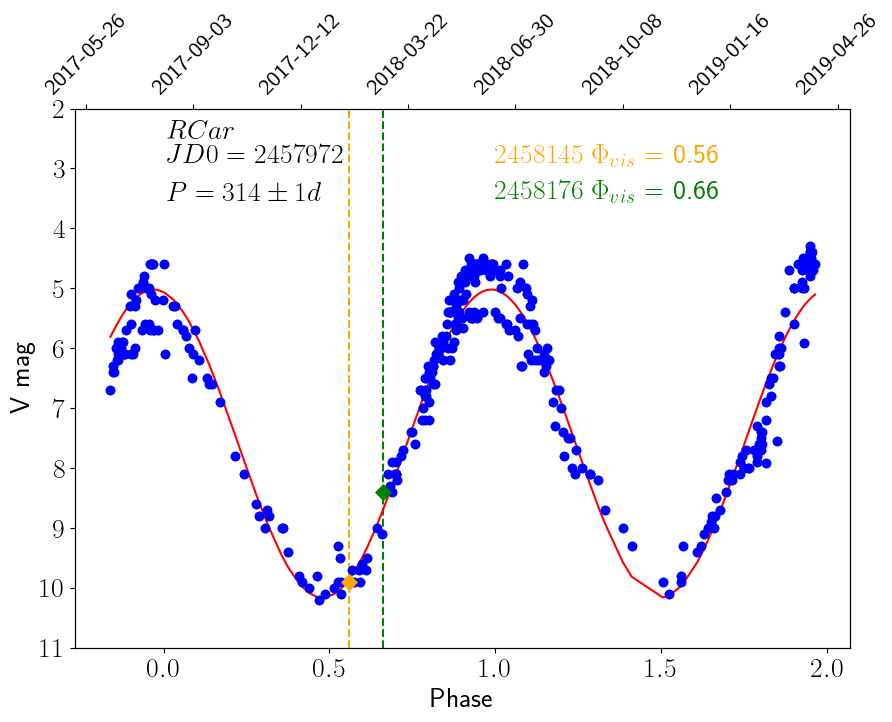}
    \caption{ Visual light curve of R Car based on the AAVSO database (blue points). The red line represents the best-fit of a sine function to the curve which results in a period of  P = 314 $\pm$ 1 days. The dashed vertical lines and the colored diamonds show the epochs of our 2018 GRAVITY observations (see the labels on the plot). }
    \label{fig:light_curve_comp}
\end{figure}

Miras are AGB stars with large-amplitude variations over long periods (10$^2$ days), for which asymmetries in the circumstellar material are quite frequent \citep{cruzalebes2015departure, ragland2006first, lacour2009pulsation}. R Car is a bright near-infrared Mira star ($K-$band: -1.23 mag) located at 182 $\pm$ 16 pc from Earth \citep{2020yCat.1350....0G}. This source exhibits variations in the V-band from +3.9 to +10.5 mag. Fig. \ref{fig:light_curve_comp} shows a visual light curve of R Car based on data obtained from the AAVSO (American Association of Variable Star Observers) database\footnote{\href{https://www.aavso.org/}{https://www.aavso.org/}}.

The photosphere of the star was observed in 2014 as part of the Interferometric Imaging Contest \citep{monnier20142014}, using the H-band 4-beam combiner interferometer PIONIER attached to the VLTI. The reconstructed image using these data revealed bright spots on its surface and (potential) inner layers of circumstellar material and a stellar disk  with a radius of 5 mas.

\subsection{GRAVITY observations and interferometric data reduction}

R Car was observed during January and February, 2018, for a total of 8 hours, with the beam-combiner GRAVITY. The data were taken using 6 baselines with the Auxiliary Telescopes (ATs) configuration A0-B2-C1-D0, which provides minimum and maximum baselines of 7.67 m (equivalent to an angular resolution $\Delta \theta \sim 59$ mas in this wavelength range) and 32 m ($\Delta \theta \sim 15 $mas),  respectively. The data cover a wavelength range between 2.0 $\mu$m and 2.4 $\mu$m with a spectral resolution of R$\sim$ 4000. The complete log of observations is reported in Table \ref{table:log_obs} in Appendix \ref{sec:observations_grav}.

We reduce and calibrate our data using the latest release of the GRAVITY pipeline (version: 1.5.0). Detailed descriptions of the recipes are available in the ESO pipeline  \href{https://www.eso.org/sci/software/pipelines/gravity/gravity-pipe-recipes.html}{manual} and in \citep{lapeyrere2014gravity}. The same standard data reduction procedures were applied to the scientific target and the calibrator star (HD\,80404; which has an angular size of 1.8 mas) to correct for the dark, flat-field and bad-pixels, as well as to extract the interferometric observables. The calibrated V$^2$ of the target were obtained following the standard procedure described in the ESO Gravity pipeline manual, which consists in dividing the raw observables by the ones of the calibrator star. Calibrated CPs were obtained by subtracting the raw CPs of the calibrator from the ones of the source. Due to the observed changes in the observables at similar baselines, the data were split into two epochs (January and February with equivalent pulsation phases of 0.56 and 0.66, respectively). The calibrated R Car data files can be downloaded directly from the Optical Interferometric Data Base (OiDB) archive hosted at the Jean-Marie Mariotti Center\footnote{\href{https://oidb.jmmc.fr/index.html}{https://oidb.jmmc.fr/index.html}}. 
\subsection{GRAVITY spectrum calibration}\label{sec:spectrum_calib}

\begin{figure}
    \centering
    \includegraphics[width=8.5cm]{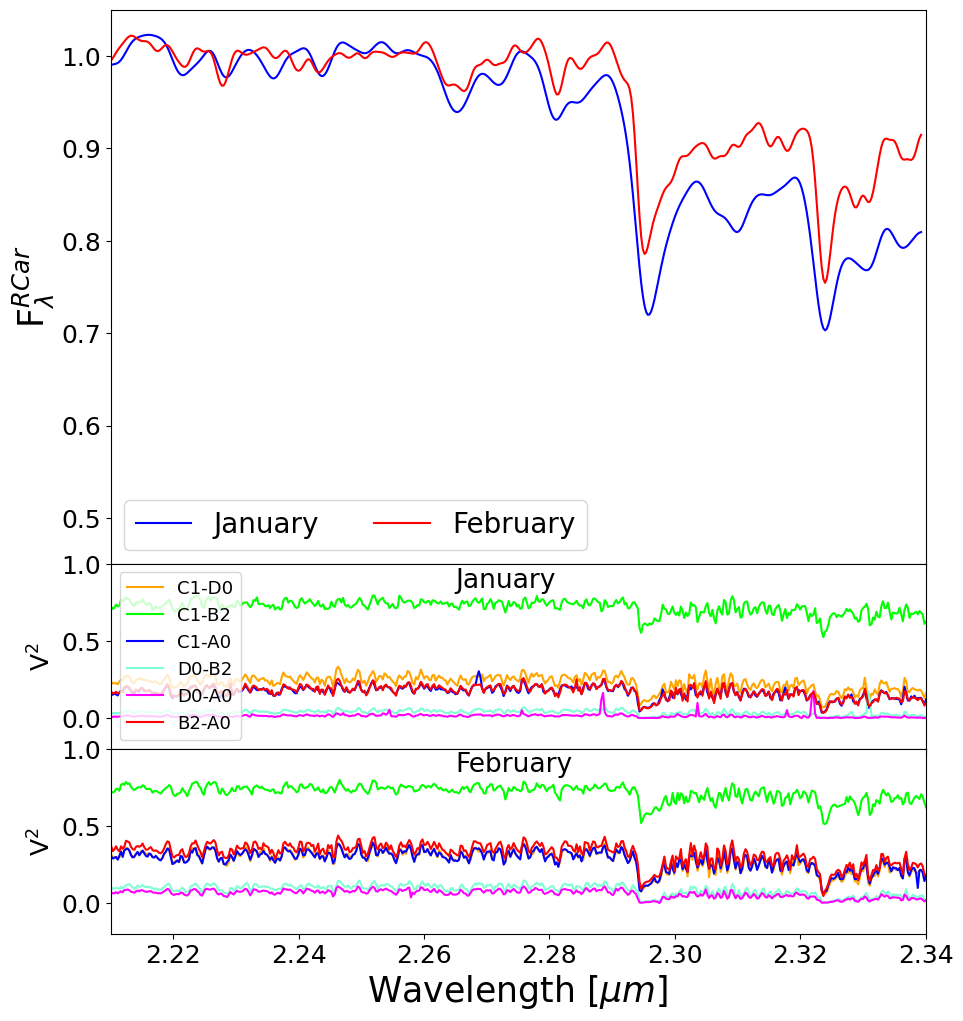}
    \caption{ The top panel shows the normalized calibrated spectra of the GRAVITY data. The blue spectrum corresponds to the January epoch while the red one corresponds to the February one (see the labels on the plot). The medium and bottom panels show plots of the V$^2$ versus wavelength for the two epochs observed. }
    \label{fig:spectra_janfeb}
\end{figure}

GRAVITY data provide us $K-$band spectra (R$\sim$4000) of the target and calibrator. For the posterior analysis of the CO bandheads observed in the R Car spectrum, we calibrate and normalize it to remove the telluric contribution from the Earth's atmosphere by using the following procedure: 

\begin{itemize}
    \item  We averaged the R Car and calibrator spectra to obtain the high signal-to-noise spectra F$_{\mathrm{avg}}^{\mathrm{R Car}}$ and F$_{\mathrm{avg}}^{\mathrm{HD\,80404}}$, respectively. 
    To remove the telluric lines from the averaged R Car spectrum, we used the following expression:
    
    \begin{equation}
        \mathrm{F}_{\lambda}^{\mathrm{R Car}} = \frac{\mathrm{F}_{\mathrm{avg}}^{\mathrm{R Car}}}{\mathrm{F}_{\mathrm{avg}}^{\mathrm{HD\,80404}}}\times {\mathrm{F}_{\mathrm{avg}}^{\mathrm{models}}}\,,
    \end{equation}
    
    where F$_{\mathrm{avg}}^{\mathrm{models}}$ was calculated by averaging 15 theoretical spectra from the Theoretical spectra web server \footnote{\href{http://svo2.cab.inta-csic.es/theory/newov2/}{http://svo2.cab.inta-csic.es/theory/newov2/}}\citep{allard2010model} of our calibrator spectral type A7\,Ib. 
    \item We flatten and normalized $F_{\lambda}^{\mathrm{R Car}}$ (in the range of 2.21 - 2.34 $\mu$m) by using a polynomial function of third order. 
    \item Finally, we applied a wavelength correction (equivalent to a Doppler velocity of 17 kms$^{-1}$) for the local standard of rest (LSR) taking into account the proper motion and the distance of the source. The normalized spectrum of R Car of each epoch is presented in Fig. \ref{fig:spectra_janfeb}.
\end{itemize}

\section{Analysis and results}\label{analysis}
\subsection{Geometrical model} \label{geo_continuum}

To obtain the angular size of R Car across the $K$-band, we applied a geometrical model of a uniform disk (UD) to the V$^2$ data. The visibility function of this model is given by the following equation \citep{berger2007introduction}:

\begin{eqnarray}\label{Eq:UD}
    V_{\mathrm{UD}}(u,v) = 2(\mathrm{F_r})\frac{J_1(\pi \rho \Theta_\mathrm{UD} )}{\pi  \rho \Theta_{\mathrm{UD}}}\,,
\end{eqnarray}

where $\rho = \sqrt{u^2+ v^2}$, $u$ and $v$ are the spatial frequencies sampled by the interferometric observations, $J_1$ is the first order Bessel function, $\Theta_\mathrm{UD}$ and F$_r$ are the angular diameter of the uniform disk profile and a scaling factor that accounts for the over-resolved flux in the observations, respectively. Prior to model the pseudo-continuum data, we average 200 spectral bins into one single measurement. This allows us to obtain eight spectral channels across the range 2.0 $\mu$m - 2.4 $\mu$m. Considering the calibration error in the averaged quantities, the final precision of the V$^2$ is $\sigma_{\mathrm{V^2}} \sim$ 0.01 and the precision of the CPs $\sigma_{\mathrm{CP}}$ is $\sim$ 1 deg. Figure \ref{vis2t3sf_uv_jan} shows the calibrated V$^2$ and CPs as well as the u-v coverage of the two epochs of observation for the eight spectral channels probed.

\begin{figure}
   \centering

    {\includegraphics[width=9cm]{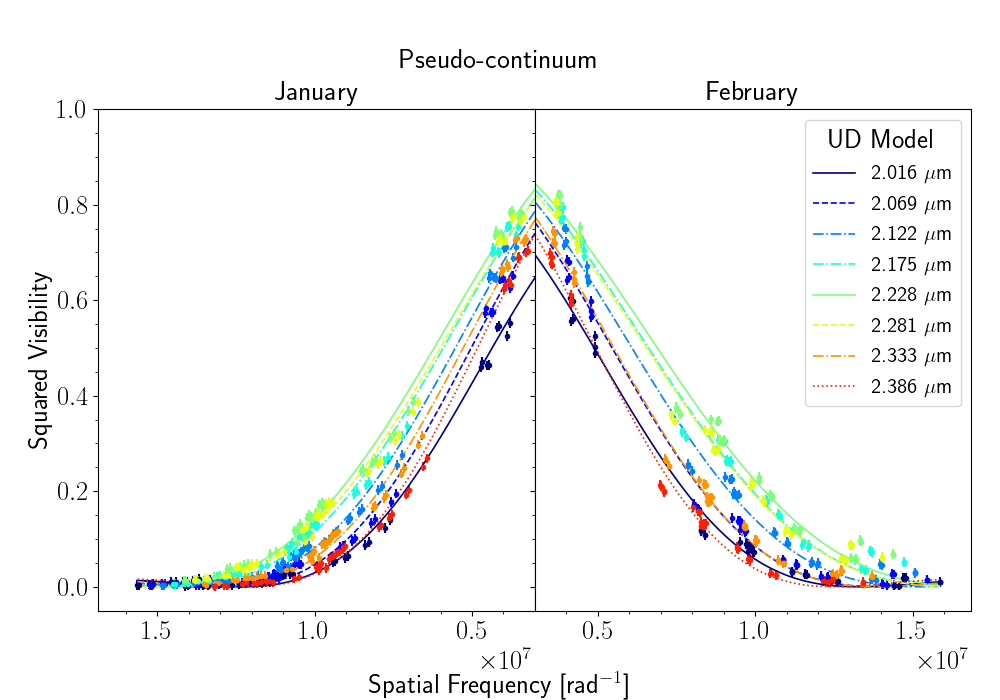}}
      \caption{V$^2$ of the pseudo-continuum versus spatial frequencies of both epochs. The lines indicate the best-fit UD models. Each color corresponds to a different wavelength (see the label on the plot). Notice that the spatial frequencies increase in opposite directions depending on the epoch. This allowed us to have a better visual comparison of the trends present in the V$^2$ data between the two epochs.
              }
         \label{all_fits2seasons}
   \end{figure}

\begin{figure}
    \centering
    \includegraphics[width=5.5cm]{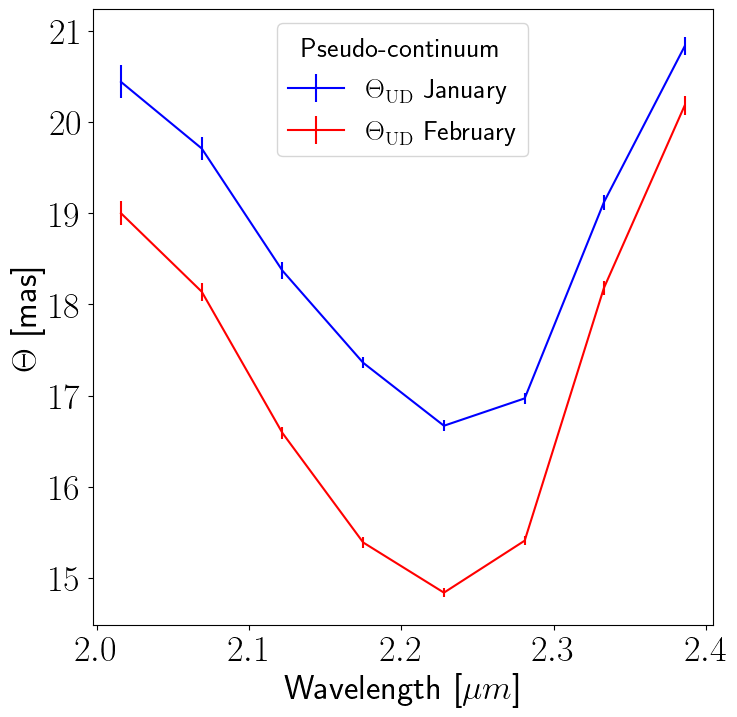}
    \caption{Best-fit $\Theta_{\mathrm{UD}}$ as a function of wavelength for the two analyzed epochs. Each color corresponds to a different epoch (see the labels on the plot).}
    \label{fig:model_comps}
\end{figure}

To fit the data, we used a Monte-Carlo Markov-Chain (MCMC) algorithm based on the \texttt{Python} package \texttt{emcee} \citep{foreman2013emcee}. We let 250 walkers evolve for 150 steps using the data of each spectral bin independently. Table \ref{table_CGresultsgeo_jan} shows the best-fit of $\Theta_{\mathrm{UD}}$ and F$_r$ for each one of the observed epochs. Figure \ref{all_fits2seasons} shows the V$^2$ data and their corresponding best-fit models. Figure \ref{fig:model_comps} shows the evolution of the best-fit $\Theta_{\mathrm{UD}}$ versus wavelength of the two epochs of observation. From this figure, we observe two significant changes in the pseudo-continuum structure of R Car. 
 
 First, the size of the target changes with wavelength, decreasing as we move towards the center of the bandpass ($\lambda \sim$ 2.23 $\mu$m) and then increasing again. This trend is expected due to the presence of molecules like H$_2$O, at $\sim$ 2 $\mu m$, and CO, at $\sim$ 2.29-2.33 $\mu m$, lying above the stellar surface \citep[see e.g.,][]{Wittkowski_2008}. The smallest $\Theta_{\mathrm{UD}}$ value (at $\sim$ 2.23 $\mu m$) corresponds to the closest pseudo-continuum emitting region due to the photosphere. This size is the best estimate of the stellar disk diameter in $K-$band. \citet{ireland2004multiwavelength} derived wavelength-dependent Gaussian FWHM between 15-20 mas at 700-920 nm. Our best-fit UD diameter at $\sim 2.23$ $\mu m$ indicates that the star appears to be larger near the V-band than in K-band. This size difference is due to the presence of dust layers obscuring the stellar-disk in the V-band, but being transparent at $K-$band, allowing us to have a closer look at the continuum emitting region. 

Second, the disk diameter is larger during the January epoch by up to 12$\%$ compared with the February epoch. From the light curve reported on Fig. \ref{fig:light_curve_comp}, it can be observed that the magnitude of the star increased in the February epoch, when the diameter of the star is smaller. This behaviour comes with an expected increment of the temperature during February, which makes the star look brighter at the cost of having a smaller radius \citep[see e.g.,][]{wittkowski2012fundamental, wittkowski2018vlti, haubois2015resolving}. A rough estimate of the temperature difference (T$_\mathrm{J}$/T$_\mathrm{F}$) between the observed epochs could be obtained using the Steffan-Boltzmann Law and the magnitude difference ($|\Delta V_{\mathrm{mag}}|$ = 1.8), from Fig.\ref{fig:light_curve_comp}, resulting in T$_\mathrm{January}$/T$_\mathrm{February} \sim$ 0.62.

\begin{table}
\caption{Best-fit parameters of the uniform disk geometrical model}       
\label{table_CGresultsgeo_jan}     
\centering                         
\begin{tabular}{c c c c c}       
\hline\hline                 
$\lambda [\mu m]$  & \multicolumn{2}{c}{$\Theta_{\mathrm{UD}}$ [mas]}  & \multicolumn{2}{c}{F$_{\mathrm{r}}$} \\
\hline                        

         & \textbf{January} & \textbf{February} & \textbf{January} & \textbf{February} \\
\hline
   2.016 & 20.44$^{+0.17}_{-0.18}$ & 19.00$^{+0.13}_{-0.13}$& 0.808$^{+0.010}_{-0.009}$ & 0.842$^{+0.009}_{-0.008}$\\[0.2cm]
    
   2.069 & 19.71$^{+0.12}_{-0.13}$ & 18.14$^{+0.09}_{-0.10}$& 0.910$^{+0.008}_{-0.008}$ & 0.908$^{+0.008}_{-0.007}$\\[0.2cm]

   2.122 & 18.37$^{+0.08}_{-0.09}$ & 16.59$^{+0.07}_{-0.07}$& 0.940$^{+0.006}_{-0.006}$ & 0.932$^{+0.006}_{-0.006}$\\[0.2cm]
    
   2.175 & 17.36$^{+0.07}_{-0.06}$ & 15.39$^{+0.06}_{-0.06}$& 0.954$^{+0.006}_{-0.006}$ & 0.942$^{+0.006}_{-0.006}$\\[0.2cm]
    
   2.228 & 16.67$^{+0.06}_{-0.06}$ & 14.84$^{+0.05}_{-0.05}$& 0.964$^{+0.005}_{-0.005}$ & 0.948$^{+0.005}_{-0.006}$\\[0.2cm]
     
   2.281 & 16.97$^{+0.06}_{-0.06}$ & 15.41$^{+0.06}_{-0.05}$& 0.945$^{+0.005}_{-0.005}$ & 0.931$^{+0.006}_{-0.005}$\\[0.2cm]
     
   2.333 & 19.12$^{+0.08}_{-0.08}$ & 18.18$^{+0.07}_{-0.08}$& 0.932$^{+0.006}_{-0.006}$ & 0.923$^{+0.007}_{-0.007}$\\[0.2cm]
    
   2.386 & 20.83$^{+0.10}_{-0.10}$ & 20.18$^{+0.10}_{-0.10}$& 0.928$^{+0.006}_{-0.007}$ & 0.915$^{+0.007}_{-0.007}$\\[0.2cm]
    
\hline                                   
\end{tabular}
\end{table}

\subsection{Spherical thin layer: the MOLsphere model}\label{sec:moslphere}

The GRAVITY spectra show prominent CO band heads (see Fig. \ref{fig:spectra_janfeb}). The data show drops in the V$^2$ of the CO 2-0 ($\lambda_0 \sim$2.2946 $\mu$m) and CO 3-1 ($ \lambda_0 \sim$2.3240 $\mu$m) vibro-rotational transitions compared with the pseudo-continuum, which indicate that the CO line-emitting region is more extended \citep[see e.g.,][]{wittkowski2012fundamental}. The innermost CO layers are important to trace the dust nucleation region since most of the carbon or oxygen present in the atmosphere is concentrated in the stable bound of this molecule \citep{hofner2018mass}. 

\begin{figure}
     \centering
     \includegraphics[width=4.5cm]{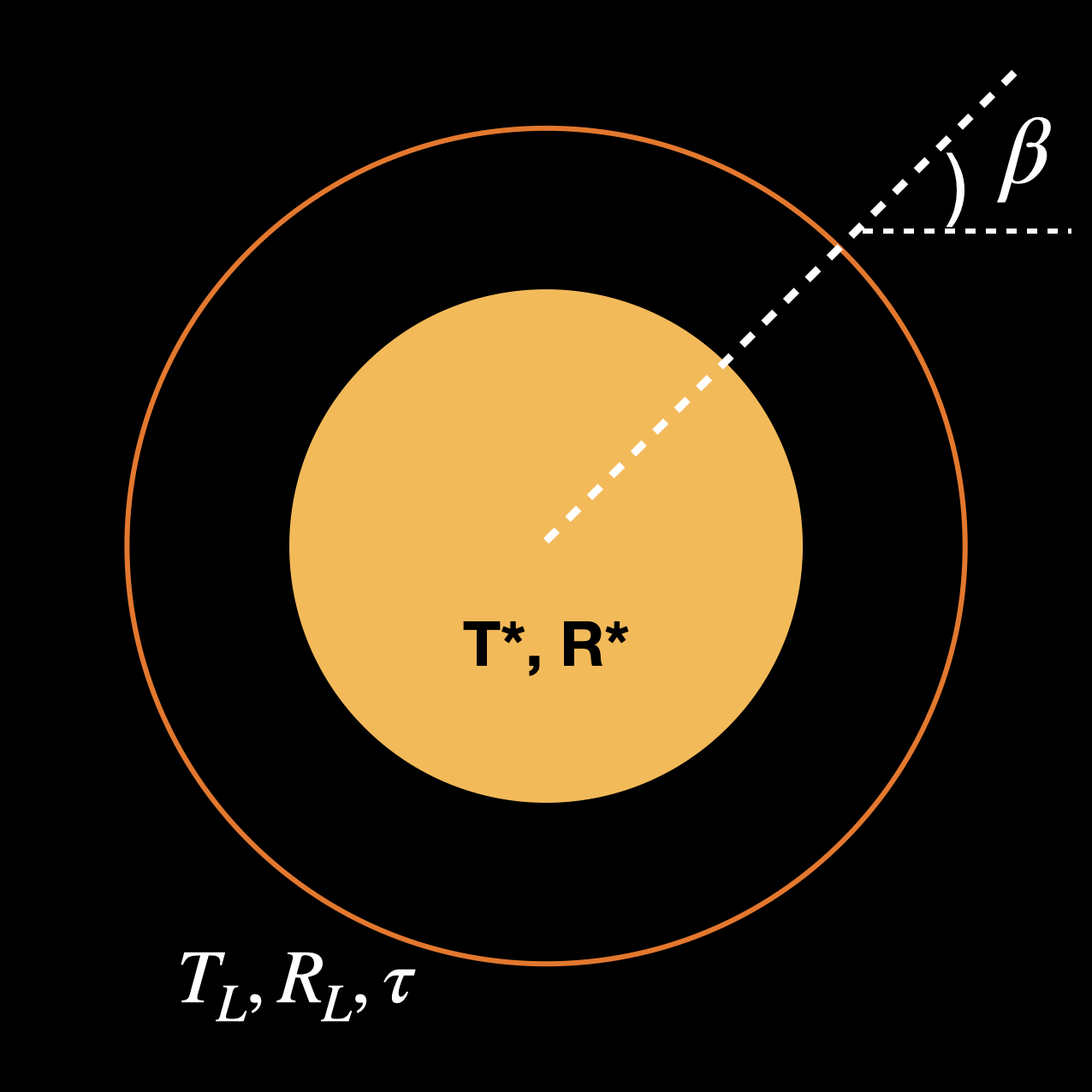}
     \caption{Illustration of the single-layer model. The yellow disk represents the star, and the orange ring represents the layer. The parameters in the image are as described in Sec. \ref{sec:moslphere}. This image was adapted from \citet{rodriguez2021molsphere}. }
     \label{fig:slmodelf}
 \end{figure}

To characterize the regions where those transitions originate, we employed a single-layer model (called MOLsphere) which has proven to be successful in explaining the absorption of the CO molecule that appears in the visibilities of evolved stars \citep{perrin2004unveiling, perrin2005study, montarges2014properties, rodriguez2021molsphere}. This model consists of a stellar disk with a compact layer around it.  The star is modeled by a stellar surface of radius $\mathrm{R_{*}}$ which emits as a black-body at a temperature $\mathrm{T_{*}}$. It is surrounded by a compact spherical layer of radius $\mathrm{R_{L}}$ that absorbs the radiation emitted by the star and re-emits it like a black-body. The MOLsphere is characterized by its temperature $\mathrm{T_{L}}$, radius $\mathrm{R_L}$ and its optical depth $\tau_{\lambda}$. The region between the stellar photosphere and the layer is assumed to be empty (see Fig. \ref{fig:slmodelf} for a sketch of the model). The analytical expression of the model is given by:

\begin{equation}
I_{\lambda}^r = \left\{ \begin{array}{lcc}
             B_\lambda(\mathrm{T_*})e^{(-\tau_\lambda/\mathrm{cos}\theta)} \\ +  B_\lambda(\mathrm{T_L})[1-e^{(-\tau_\lambda/\mathrm{cos}\theta)}] &if& \mathrm{r} \leq \mathrm{R_*} \\
             \\ B_\lambda(\mathrm{T_L})[1-e^{(-2\tau_\lambda/\mathrm{cos}\theta)}] &if& \mathrm{R_*} < \mathrm{r} \leq \mathrm{R_L} \\
             \\ 0 & & \mathrm{otherwise},
             \end{array}
   \right.
   \label{layer_star}
\end{equation}

\noindent where $I_{\lambda}^r = I_{\lambda}^r(\mathrm{T_{*}}, \mathrm{T_L}, \mathrm{R_*}, \mathrm{R_L}, \tau_{\lambda})$,  $\mathrm{T_*}$ and $\mathrm{T_L}$ are the temperatures of the photosphere and of the CO layer, respectively; $\mathrm{R_*}$ and $\mathrm{R_L}$ are the angular radius of the star and the layer, respectively; $\tau_\lambda$ is the optical depth of the molecular layer at wavelength $\lambda$; $B_\lambda(\mathrm{T})$ is the Planck function (at wavelength $\lambda$ and temperature T); and $\beta$ is the angle between the radius vector and the line-of-sight so that $\cos\beta = \sqrt{1-(\mathrm{r}/\mathrm{R_L})^2}$.

 For the fitting, we adopted a $\mathrm{T_{*}}$=2800 K and $\mathrm{R_{*}}$=5 mas \citep{monnier20142014,mcdonald2012fundamental}. We use the disk estimation reported in the $H-$band because it is considerably less affected from molecular contribution, in contrast with the K-band where  molecules like CO, H$_2$O, and OH among others \citep[see e.g.,][]{wittkowski2018vlti, paladini2011interferometry} are present over the entire band. The unknown parameters are then $\mathrm{T_L}$, $\mathrm{R_L}$ and $\tau_L$. We performed a two-step process to estimate these parameters. As first step, for each pair of $\mathrm{T_L}$ and $\mathrm{R_L}$, we performed a least-squares minimization to find the best-fit value of $\tau_L$ that reproduces the spectrum $F_{\lambda}^{\mathrm{R Car}}$. The next step consisted of using an MCMC method based on the \texttt{Python} library \texttt{emcee} to estimate the best combinations of $\mathrm{T_L}$, $\mathrm{R_L}$, and $\tau_L$ that reproduce the V$^2$. 
 
\begin{figure*}
    \centering
    \includegraphics[width=14cm]{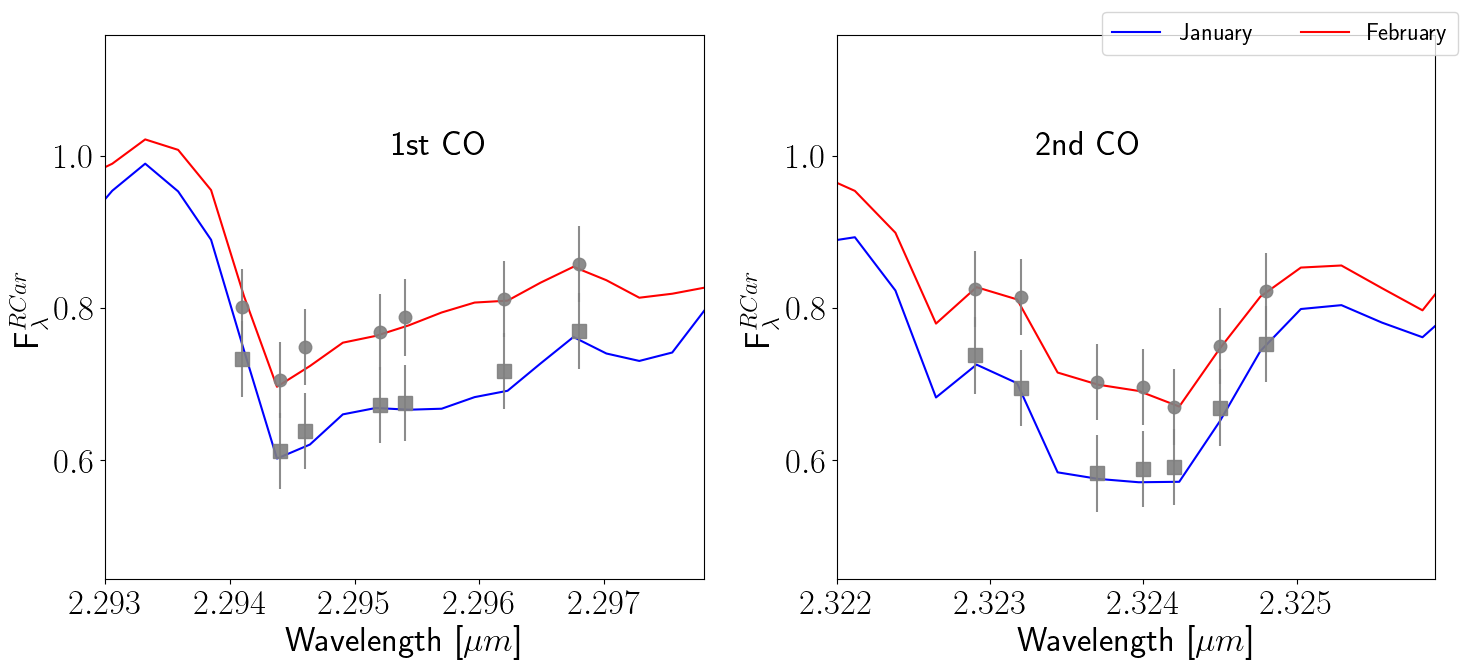}
    \caption{Best-fit to the spectra with the single-layer model for the two CO band heads and epochs. The gray dots correspond to the data points and the solid-lines correspond to the best-fit model (see the labels on the plot). }
    \label{fig:SLMallCO}
\end{figure*}

To account for the over-resolved flux when modeling the V$^2$ data, we added a scaling factor Fr. This value was estimated simultaneously with the other parameters in the model. Table \ref{tab:TLM_jan_firstCO} shows the corresponding results for the good quality spectral channels across the two CO band heads. Figure \ref{fig:SLMallCO} shows the R Car's spectra with the corresponding flux values obtained with the single-layer model. From the best-fit models, we derive a temperature range between 1300 - 1500 K, which is consistent with the expected temperature values to produce the CO vibro-rotational transitions observed in $K-$band \citep{geballe2007infrared}. The values of the optical depth, $\tau_L$, support that the layer is optically thick across all the spectral channels, with higher $\tau_L$ towards the centers of the band heads. In section \ref{sec:discussion}, we discuss the implications of these results in the context of the wind-driving dust candidates that can coexist with the CO.

\begin{figure*}
    \centering
    \includegraphics[width=11.5cm]{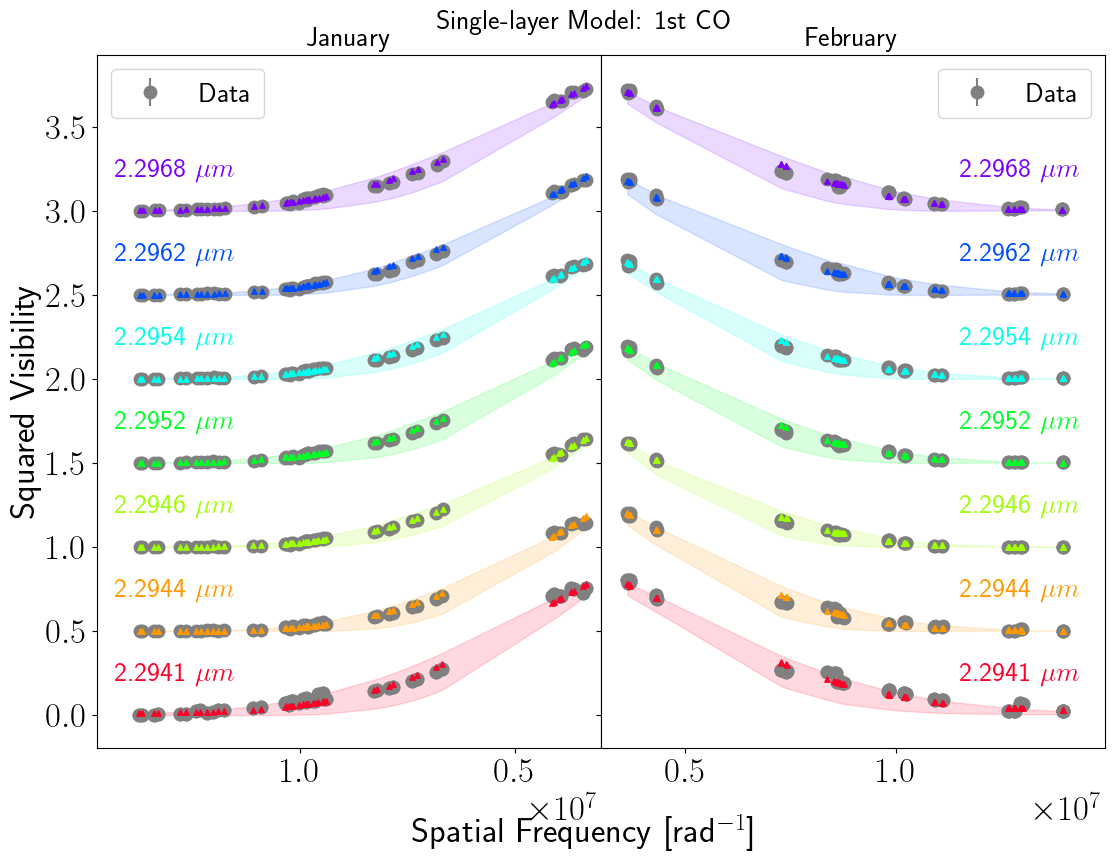}
    \caption{ Best-fit to the V$^2$ from the best single-layer models across the 1st CO band head for our two epochs. The colored dots and the shaded regions correspond to the best-fit models and their corresponding error-bars. Each color correspond to a different wavelength (see labels on the plot). The data are shown with black dots. Notice that the spatial frequencies increase in opposite directions depending on the epoch. This allowed us to have a better visual comparison of the trends present in the V$^2$ data between the two epochs. Also, for a better visualization, we displaced vertically the V$^2$.}
    \label{fig:best_SLM_1stCO}
\end{figure*}
\begin{figure*}
    \centering
    \includegraphics[width=11.5cm]{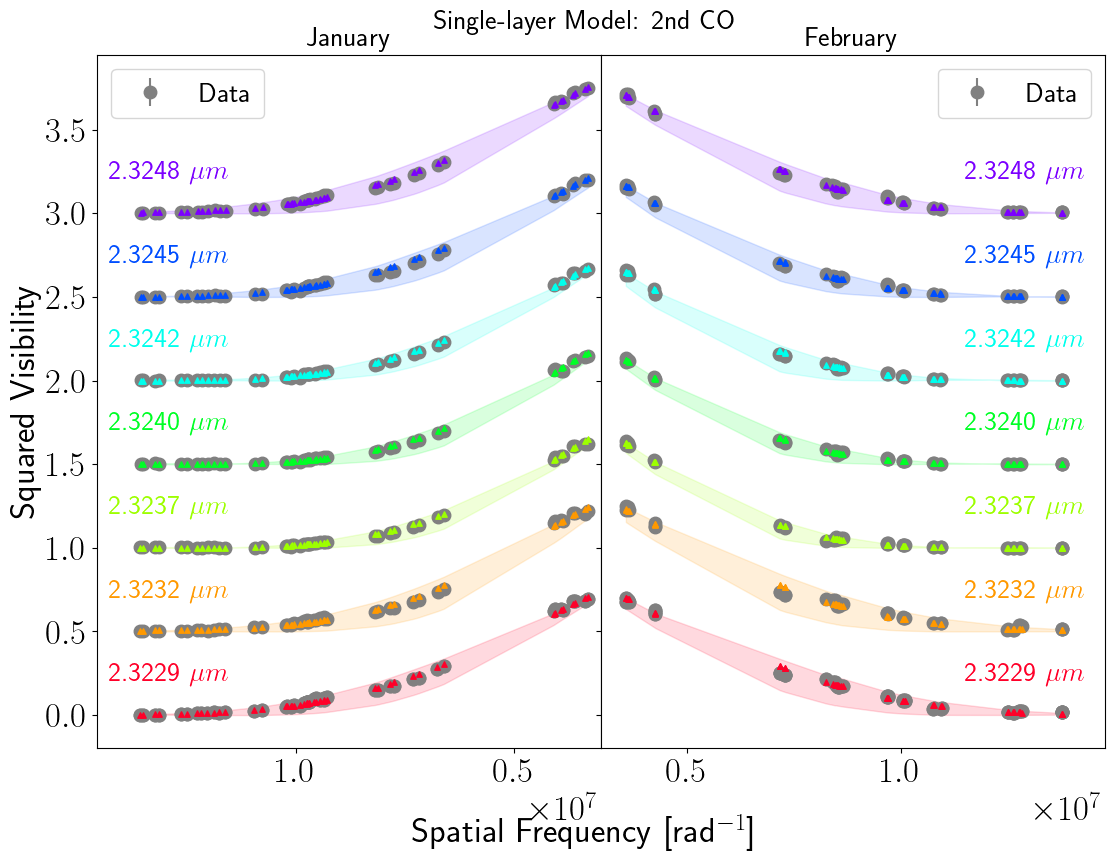}
    \caption{Best-fit to the V$^2$ from the best single-layer models across the 2nd CO band head for our two epochs. The panels are as described in Fig. \ref{fig:best_SLM_1stCO}.}
    \label{fig:best_SLM_2ndCO}
\end{figure*}

\begin{table*}
\centering
\caption{Parameters of the single-layer model for the two CO band heads and our two epochs}
\begin{tabular}{c|cccc|cccc}
\hline\hline
$\lambda [\mu m]$ & \multicolumn{1}{c}{$T_L$ [K]} & \multicolumn{1}{c}{$R_L$ [mas]}&\multicolumn{1}{c}{$\tau_L$} & Fr & \multicolumn{1}{c}{$T_L$ [K]} & \multicolumn{1}{c}{$R_L$ [mas]}&\multicolumn{1}{c}{$\tau_L$} & \multicolumn{1}{c}{Fr}\\
\hline 
 &\multicolumn{4}{c|}{\textbf{January}} & \multicolumn{4}{c}{\textbf{February}} \\[0.2cm]
\hline
\multicolumn{9}{c}{CO 2-0}\\[0.2cm]
\hline
2.2941 & 1316$^{+ 136}_{-73}$ & 13.02$^{+1.33}_{-1.73}$  & 1.3$^{+0.3}_{-0.6}$& 0.968$^{+0.029}_{-0.030}$ & 1320$^{+ 94}_{-60}$   & 12.61$^{+1.09}_{-1.27}$    & 0.9$^{+0.2}_{-0.5}$& 0.973$^{+0.024}_{-0.018}$ \\ [0.2cm]
2.2944 & 1344 $^{+90}_{-105}$  & 12.41 $^{+1.65}_{-1.05}$ & 1.9 $^{+0.5}_{-0.8}$ & 0.976$^{+0.028}_{-0.029}$ & 1366 $^{+120}_{-90}$  & 12.38 $^{+1.35}_{-1.28}$ & 1.6 $^{+0.5}_{-0.6}$ & 1.005 $^{+0.033}_{-0.024}$                   \\[0.2cm]
2.2946 & 1360 $^{+106}_{-90}$ & 12.17 $^{+1.22}_{-1.21}$ & 1.9 $^{+0.6} _{-0.9}$      & 0.944 $^{+0.027}_{-0.029}$ & 1413 $^{+103}_{-117}$ & 12.12 $^{+1.60}_{-1.09}$ & 1.7 $^{+0.6} _{-0.8}$      & 0.956 $^{+0.031}_{-0.032}$               \\[0.2cm]
2.2952 & 1358 $^{+128}_{-84}$  & 12.22 $^{+1.25}_{-1.34}$ & 1.6 $^{+0.5} _{-0.8}$       & 0.981 $^{+0.030}_{-0.026}$ & 1413 $^{+129}_{-96}$  & 11.95 $^{+1.26}_{-1.24}$ & 1.5 $^{+0.6} _{-0.8}$       & 0.981 $^{+0.025}_{-0.019}$                \\[0.2cm]
2.2954 & 1389 $^{+95}_{-83}$  & 11.90 $^{+1.15}_{-1.02}$ & 1.7 $^{+0.5} _{-0.6}$       & 0.976 $^{+0.031}_{-0.026}$  &  1406 $^{+108}_{-83}$  & 12.07 $^{+1.26}_{-1.03}$ & 1.4 $^{+0.4} _{-0.6}$       & 0.989 $^{+0.028}_{-0.022}$              \\[0.2cm]
2.2962 & 1448 $^{+92}_{-111}$  & 11.28 $^{+1.23}_{-0.83}$ & 1.7 $^{+0.6} _{-0.9}$       & 0.962 $^{+0.026}_{-0.025}$  & 1394 $^{+195}_{-88}$  & 12.23 $^{+1.39}_{-1.78}$ & 1.2 $^{+0.4} _{-0.7}$       & 0.976 $^{+0.029}_{-0.033}$               \\[0.2cm]
2.2968 & 1424 $^{+148}_{-93}$  & 11.69 $^{+1.17}_{-1.30}$  & 1.4 $^{+0.4} _{-0.7}$ & 0.987 $^{+0.023}_{-0.024}$ & 1504 $^{+171}_{-126}$  & 10.95 $^{+1.44}_{-1.18}$  & 1.3 $^{+0.4} _{-0.7}$ & 0.972 $^{+0.022}_{-0.023}$     \\[0.2cm]
\hline
\multicolumn{9}{c}{CO 3-1}\\[0.2cm]
\hline
2.3229 & 1420 $^{+123}_{-94}$ & 11.41$^{+1.19}_{-1.05}$& 1.5$^{+0.5}_{-0.7}$ &0.958 $^{+0.026}_{-0.022}$ &1467$^{+159}_{-119}$ & 11.04$^{+1.33}_{-1.20}$& 1.2$^{+0.4}_{-0.7}$& 0.960 $^{+0.021}_{-0.025}$  \\ [0.2cm]
2.3232 & 1316 $^{+131}_{-75}$  & 12.90 $^{+1.43}_{-1.45}$ & 1.5 $^{+0.3}_{-0.6}$ & 1.010 $^{+0.027}_{-0.030}$ &1406 $^{+152}_{-100}$  & 11.79 $^{+1.42}_{-1.40}$ & 1.2 $^{+0.3}_{-0.6}$ & 0.996 $^{+0.024}_{-0.025}$        \\[0.2cm]
2.3237 & 1309 $^{+82}_{-76}$ & 12.78 $^{+1.20}_{-1.07}$ & 2.1 $^{+0.5} _{-0.7}$  & 0.958 $^{+0.032}_{-0.033}$ &1336 $^{+86}_{-73}$ & 13.33 $^{+1.18}_{-1.09}$ & 1.9 $^{+0.5} _{-0.7}$      & 0.990 $^{+0.028}_{-0.026}$              \\[0.2cm]
2.3240 & 1315 $^{+81}_{-86}$  & 12.67 $^{+1.30}_{-1.09}$ & 2.1 $^{+0.6} _{-0.8}$ & 0.969 $^{+0.030}_{-0.035}$ &1351 $^{+102}_{-79}$  & 12.80 $^{+1.31}_{-1.15}$ & 1.8 $^{+0.5} _{-0.7}$       & 0.971 $^{+0.034}_{-0.027}$    \\[0.2cm]
2.3242 & 1330 $^{+91}_{-81}$  & 12.18 $^{+1.15}_{-1.07}$ & 1.9 $^{+0.6} _{-0.8}$ & 0.960 $^{+0.027}_{-0.027}$ &1352 $^{+116}_{-110}$  & 12.55 $^{+1.88}_{-1.29}$ & 1.8 $^{+0.5} _{-0.9}$       & 0.981 $^{+0.034}_{-0.037}$  \\[0.2cm]
2.3245 & 1394 $^{+112}_{-113}$ & 11.52 $^{+1.51}_{-1.07}$ & 1.7 $^{+0.6} _{-0.7}$  & 0.964 $^{+0.026}_{-0.025}$ &1401 $^{+122}_{-106}$  & 12.02 $^{+1.39}_{-1.17}$ & 1.5 $^{+0.5} _{-0.8}$       & 0.969 $^{+0.024}_{-0.024}$  \\[0.2cm]
2.3248 & 1393 $^{+132}_{-88}$  & 11.83 $^{+1.24}_{-1.26}$  & 1.4 $^{+0.4} _{-0.6}$ & 0.995 $^{+0.025}_{-0.024}$ &1452 $^{+142}_{-104}$  & 11.45 $^{+1.25}_{-1.15}$  & 1.3 $^{+0.4} _{-0.6}$        & 0.979 $^{+0.021}_{-0.022}$ \\[0.2cm]
\hline
\end{tabular}
\label{tab:TLM_jan_firstCO}
\end{table*} 

\subsection{Image reconstruction}\label{sec:img_reconstruction}

\subsubsection{Regularized reconstructed images}\label{subsec:reconstruct_imgs}

The geometrical model described in Sec. \ref{geo_continuum} and the MOLsphere model in Sec. \ref{sec:moslphere} considered that the source is symmetric. However, the CPs observed in our data {differ} considerably from zero or 180 degrees, which indicate that the source is asymmetric (see Fig. \ref{vis2t3sf_uv_jan} in Appendix \ref{sec:observations_grav}). So, in order to depict the complex nature of the source across the K-band and at the CO band heads, we reconstruct aperture-synthesis images using the software BSMEM \citep{baron2008image}. This code uses an iterative regularized minimization algorithm of the form:

\begin{equation}
    f(\mathbf{x}) = f_{\mathrm{data}}(\mathbf{x}) + \mu f_{\mathrm{prior}}(\mathbf{x})\,,
\end{equation}

\noindent where $f_{data}(\mathbf{x})$ is a measurement of the log-likelihood and it characterizes the discrepancy between the image model (at a given iteration) and the observables; $f_{prior}(\mathbf{x})$ contains the \textit{prior} information included in the reconstruction, and the hyperparameter $\mu$ is used to adjust the relative weight of the constraints between the measurements and the ones set by the priors \citep{renard2011image}. BSMEM uses Entropy \citep{gull1984max} as regularizer in $f_{prior}(\mathbf{x})$. 

As initial setup for our reconstruction, we created images of 128 $\times$ 128 pixels with a pixel scale of 0.469 mas. The best-fit sizes of the UD models were used as starting point in the reconstruction process. We perform the reconstructions by taking all the V$^2$ and CPs on each individual channel, separately. Since our data are constrained by a relatively sparse u-v coverage, this could affect the quality of the reconstructed images. Therefore, to properly distinguish intrinsic astrophysical features from image reconstruction artifacts \citep{haubois2015resolving, sanchez2018chromatic} and; to provide an estimate on the quality of our reconstructed images, we employed a bootstrapping method \citep{zoubir2004bootstrap}. 

We build new data samples from the original one. The samples were created by keeping the original number of data points unaltered, but allowing random sampling with replacement. This created data sets with points with higher weights than in the original data set and, some others with zero weights. This algorithm also produces slight changes in the u-v plane, which allows us to trace the impact of the u-v coverage on the reconstructed images, without loosing the statistical significance of the original number of data points. According to \cite{babu1983inference}, the statistical moments such as mean, variance and standard deviation of the bootstrapped samples are good approximations to the statistical moments of the original data sets.

\begin{figure*}
  \centering
   \subfigure[]{\includegraphics[width=16.3cm]{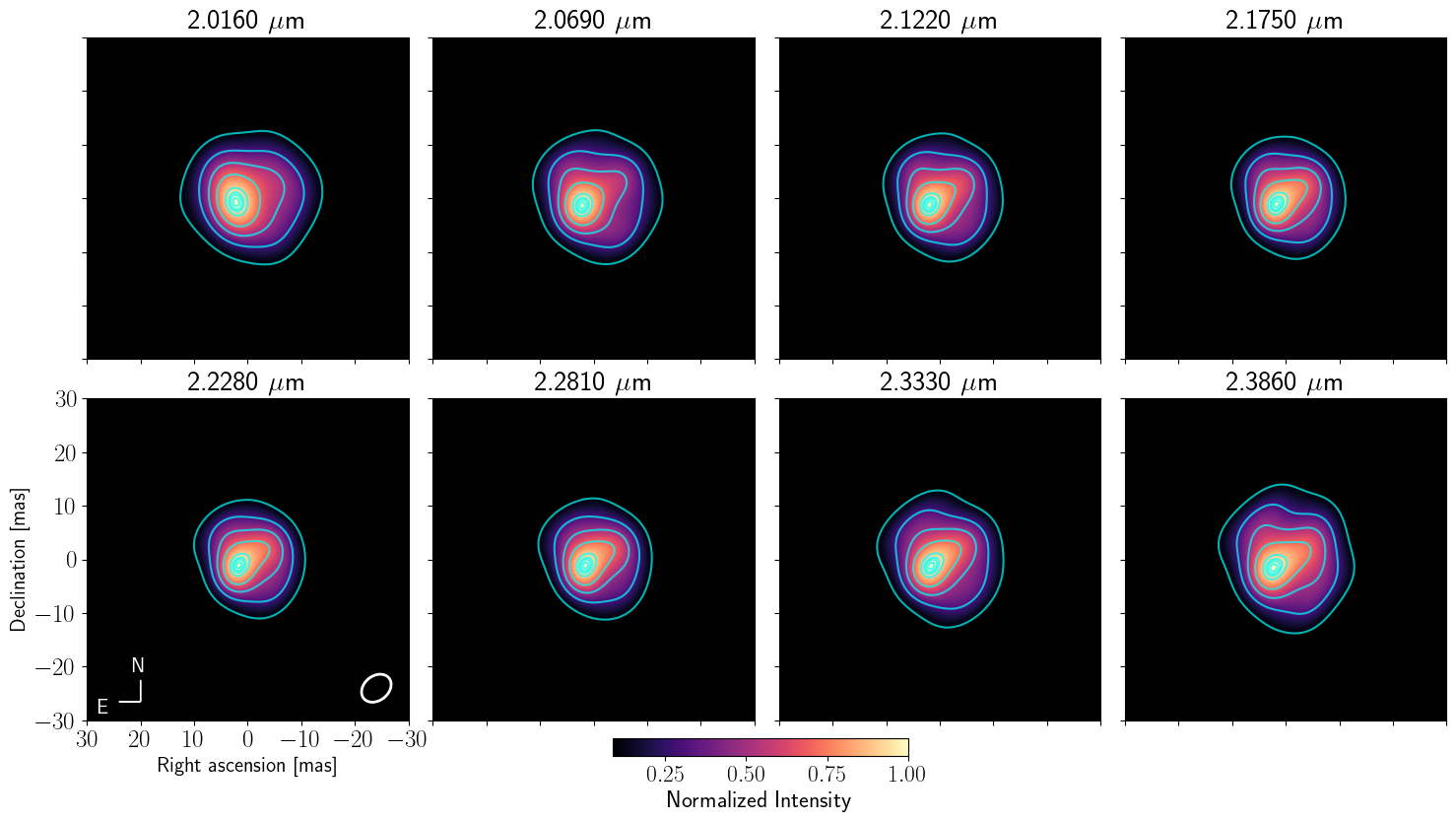}}
   \subfigure[]{\includegraphics[width=16.3cm]{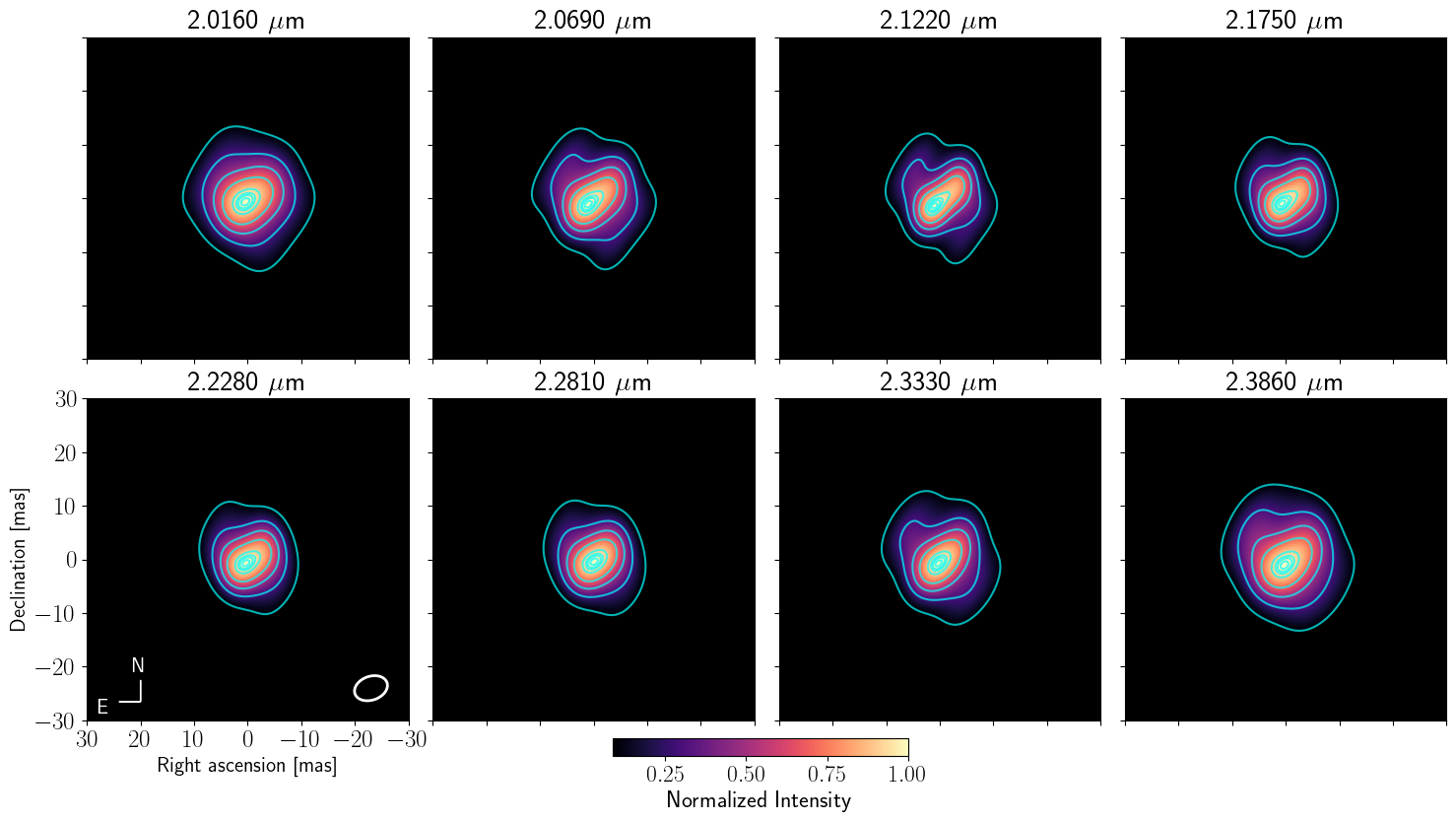}}

      \caption{Best pseudo-continuum reconstructed images for (a) the January and (b) February epochs. The white ellipse located in the right corner of the image at 2.0160 $\mu m$ corresponds to the mean synthesized primary beam. The blue contours represent the 10, 30, 50, 70 90, 95, 97 and 99 $\%$ of the intensity's peak.}
         \label{fig:beauty_janfeb_all}
   \end{figure*}

\begin{figure*}
    \centering
    \subfigure[]{\includegraphics[width = 16.3cm]{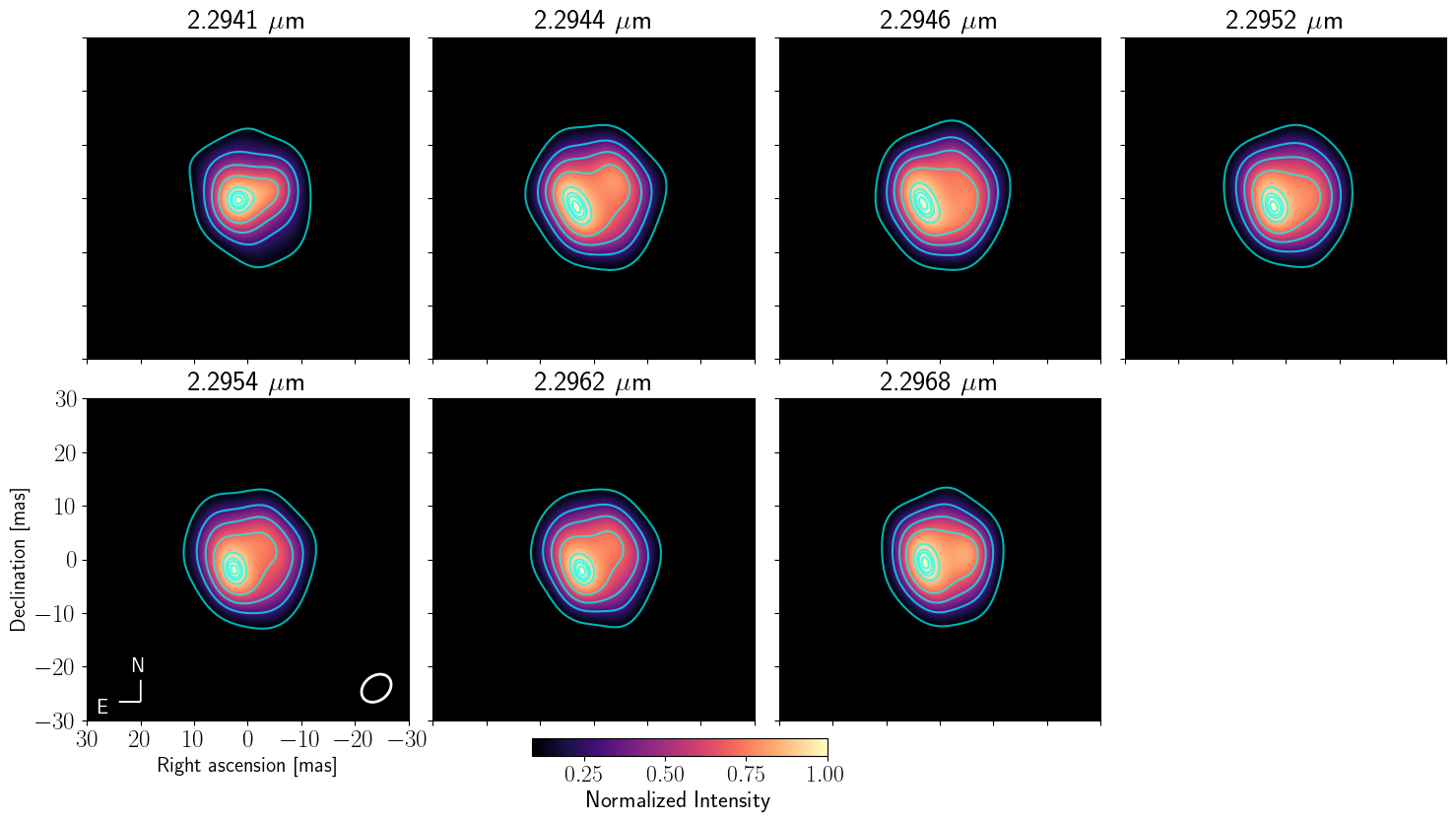}}
    \subfigure[]{\includegraphics[width = 16.3cm]{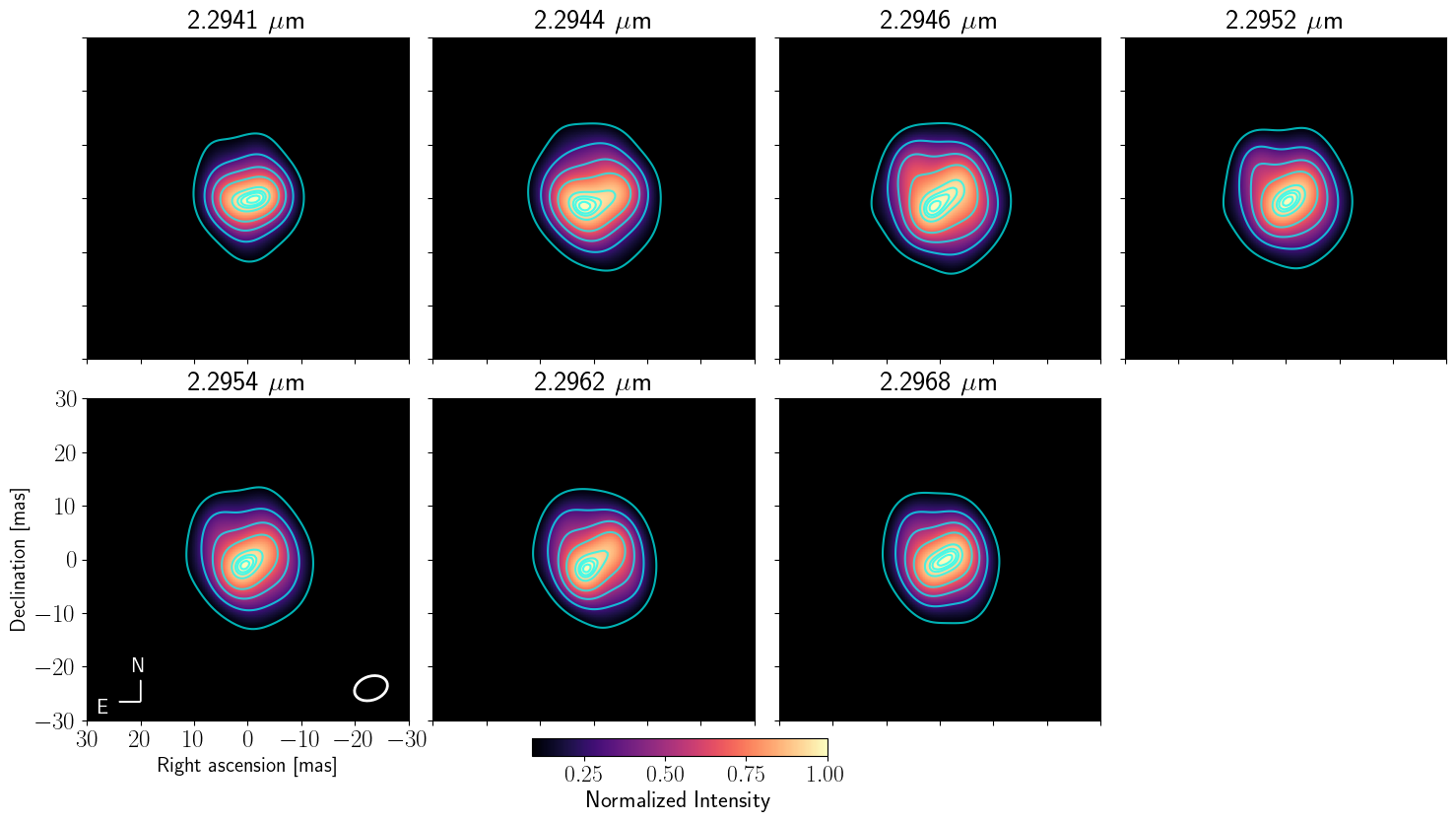}}
    \caption{Best reconstructed images per spectral channel for (a) the January and (b) the February epochs across the 1st CO band head. The white ellipse located at the right corner of the image at 2.2941 $\mu m$, and the blue contours are as described in Figure \ref{fig:beauty_janfeb_all}}
    \label{fig:bty1COimg_janfeb}
\end{figure*}

\begin{figure*}
    \centering
    \subfigure[]{\includegraphics[width = 16.3cm]{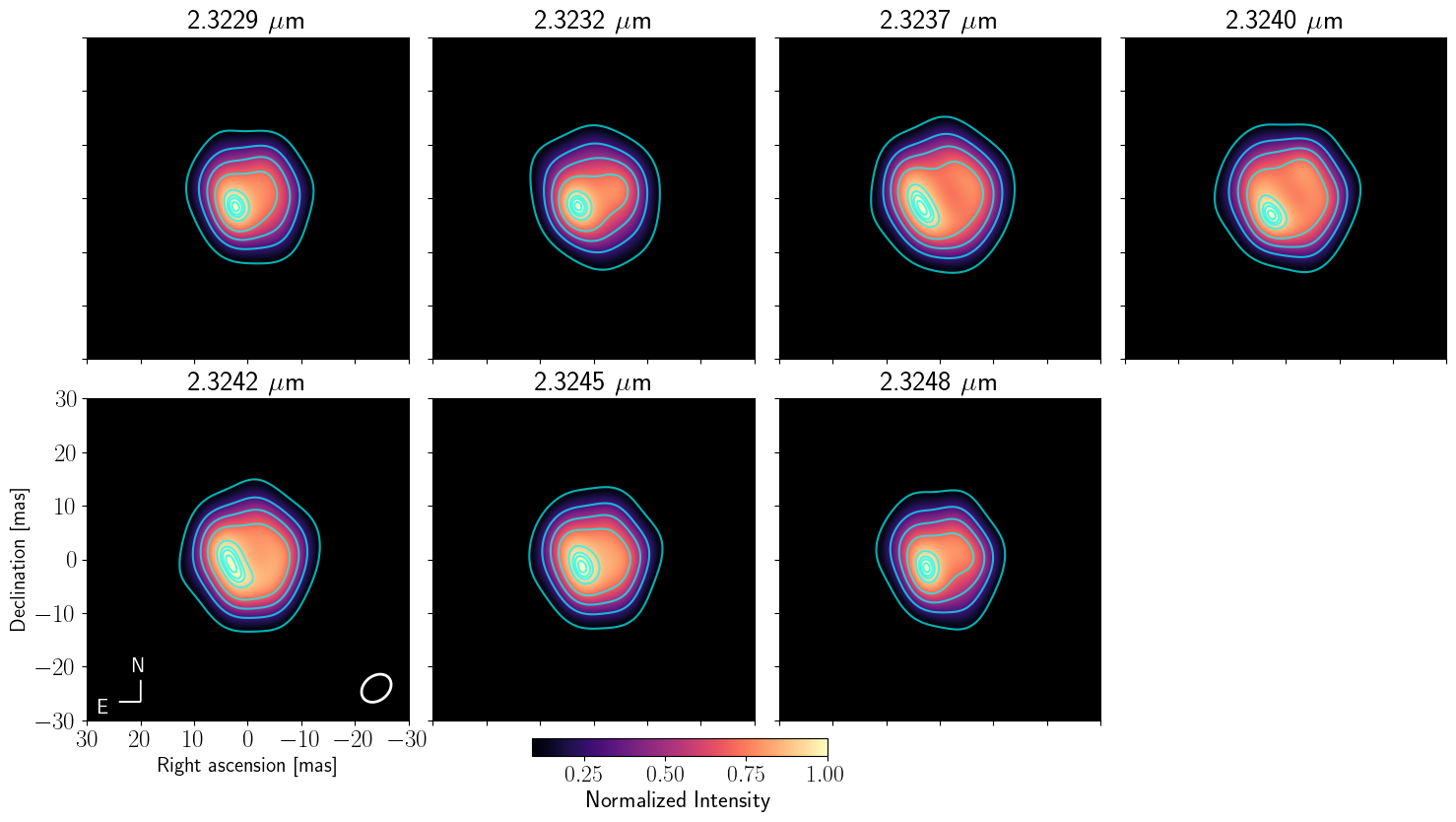}}
    \subfigure[]{\includegraphics[width = 16.3cm]{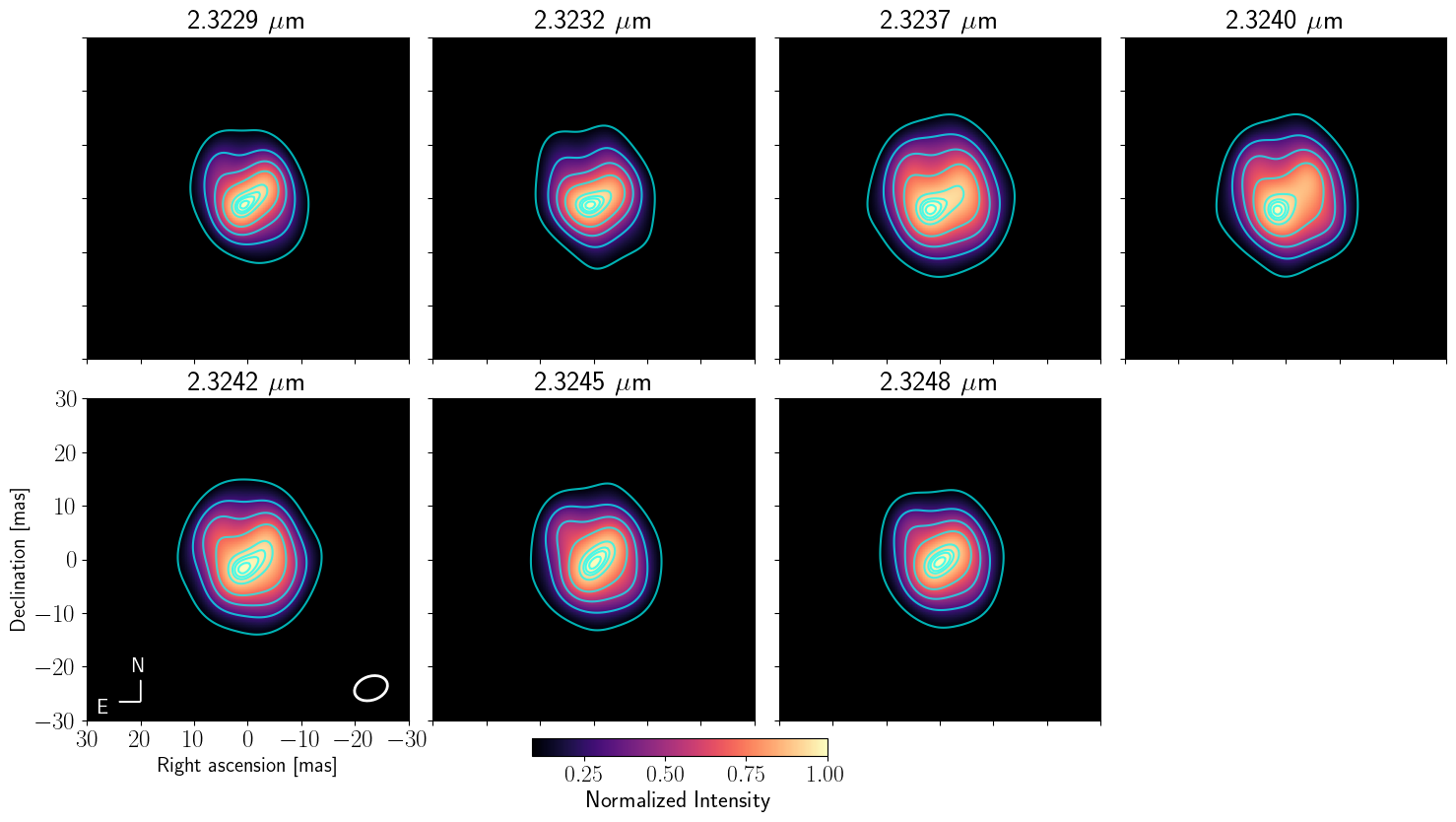}}
    \caption{Best reconstructed images per spectral channel for (a) the January, and (b) the February epoch across the 2nd CO band head. The white ellipse located  at the right corner of the image at 2.3229 $\mu m$ and the blue contours are as described in Figure \ref{fig:beauty_janfeb_all}}
    \label{fig:bty2COimg_janfeb}
\end{figure*}

We employed this method by creating 50 different sampled data sets from the original observables, $\mathbf{V}^2* = {\mathbf{V}^2_1, ..., \mathbf{V}^2_{50}}$ and $\mathbf{CP}* = {\mathbf{CP}_1, ...,\mathbf{CP}_{50}}$, using the \texttt{Python} \texttt{scikit learn} package \citep{pedregosa2011scikit}. We performed image reconstructions on each sample to obtain a set of bootstrapped images $I* = {I_1 ..., I_{50}}$. All the reconstructions from the samples converged and the best-fit image reproduces quite well the trend in the sampled observables (V$^2$ and CPs). We calculated the mean and standard deviation per pixel along the 50 bootstrapped images to obtain a mean image $I_{\mathrm{mean}}$ and a standard deviation map $I_\sigma$. We used $I_{\mathrm{mean}}/I_\sigma$ to estimate the signal to noise ratio (SNR) of the structures present in the mean image. We discarded all the features with a SNR < 3 as they are related to reconstruction artefacts. Figs. \ref{fig:beauty_janfeb_all} - \ref{fig:bty2COimg_janfeb} show the $I_{mean}$ images for the pseudo-continuum and CO band head channels recovered. All the observed structures in the $I_{\mathrm{mean}}$ images lie within the three SNR contour. Figs. \ref{fig:vis2_obs_jan_cont} to \ref{fig:vis2_obs_febm_cont} in Appendix \ref{sec:observations_grav} show the best-fit synthetic observables obtained from the  $I_{\mathrm{mean}}$ images to the original data points. As it can be seen, the trends are well-reproduced at all angular scales and wavelengths.

From a visual inspection of our best-reconstructed images, $I_{mean}$ (Figs. \ref{fig:beauty_janfeb_all} - \ref{fig:bty2COimg_janfeb}), we note that all the pseudo-continuum images are asymmetric and show structural differences between the observed epochs. In our January images we observe a spot towards the south-east direction; while in February, a more elongated (and centered) structure is observed. Excepting the change in size across the band-pass, we do not see significant structural changes with wavelength. 
 
\subsubsection{Principal Component Analysis to understand the structure of R Car}\label{subsec:PCA}

To have a more precise characterization of the asymmetries in the reconstructed images of the CO band-heads, we used the Principal Component Analysis (PCA) described by \citet{medeiros2018principal}. Those authors demonstrate that the visibilities of the Principal Components are equal to the Principal Components of the visibilities. This method is useful to trace the changes across a set of images that have the largest effect on the visibility (amplitude and -closure- phase) profile. In our case, we estimate the most significant structural changes of the observed asymmetric structures across wavelength for each of the observed epochs. The following procedure was applied to the ensemble of wavelength dependent images per epoch to extract their Principal Components.

For the PCA analysis, we normalize each data set by subtracting the corresponding mean image and dividing by their standard deviation image across the wavelength range. To perform the PCA analysis, we used the \texttt{CASSINI-PCA}\footnote{\href{https://github.com/cosmosz5/CASSINI}{https://github.com/cosmosz5/CASSINI}} package. This software allows to compute the covariance matrix of the data set and to transform it into the space of the principal components. This allows us to determine the eigenvectors (or eigen-images, in our case) and their corresponding eigenvalues. Since we only have seven images per data set,  and the possible number of components must be smaller or equal than the number of images in the data set, we decided to only keep the first four principal components. From our tests, we observed that those components explain, at least, the 93\% of the variance in the data sets. 

\begin{figure*}
    \centering
    \subfigure[]{\includegraphics[width=9cm]{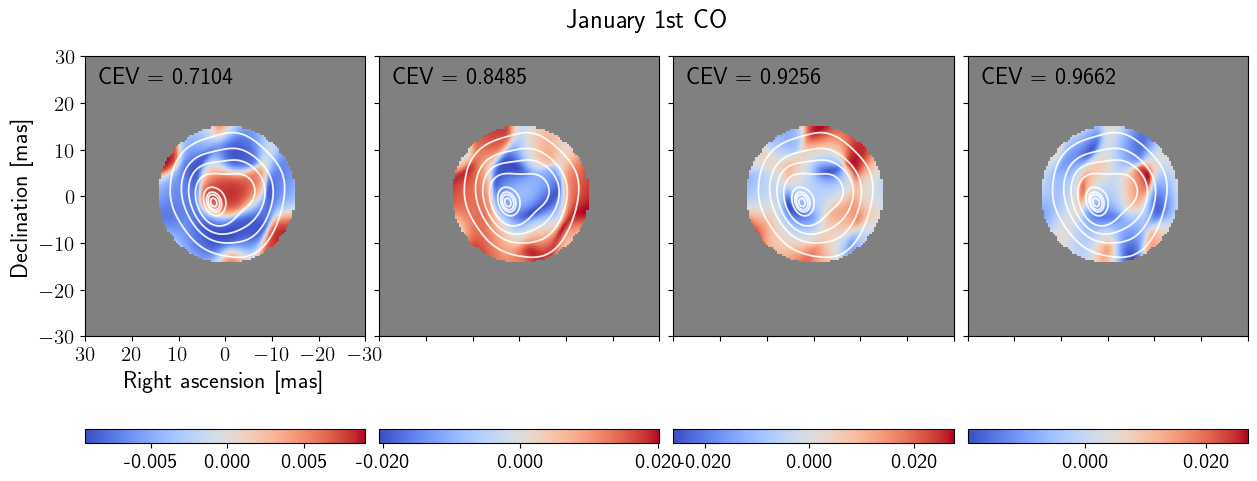}}
    \subfigure[]{\includegraphics[width=9cm]{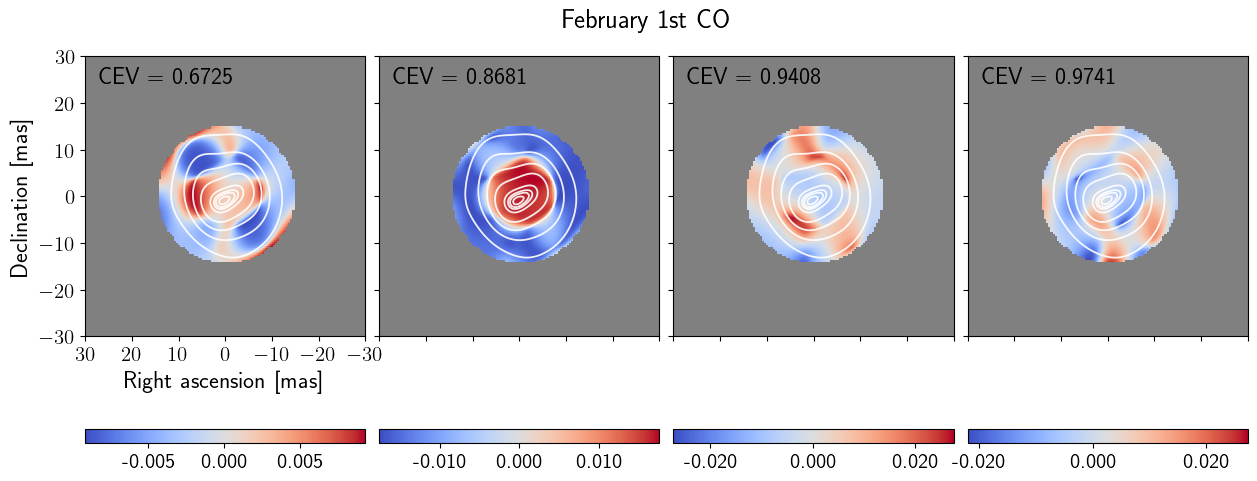}}
    \subfigure[]{\includegraphics[width=9cm]{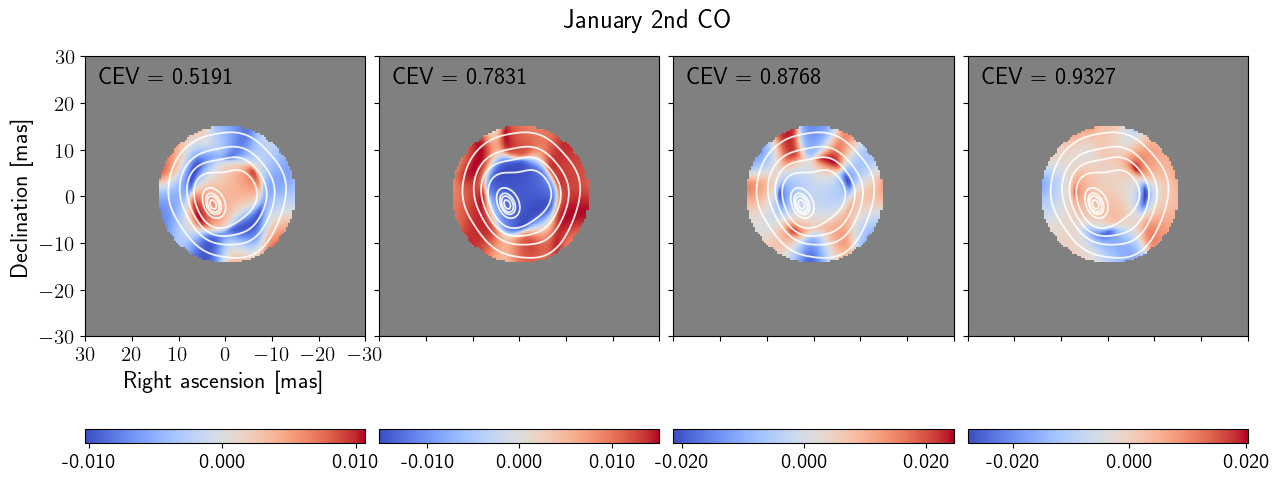}}
     \subfigure[]{\includegraphics[width=9cm]{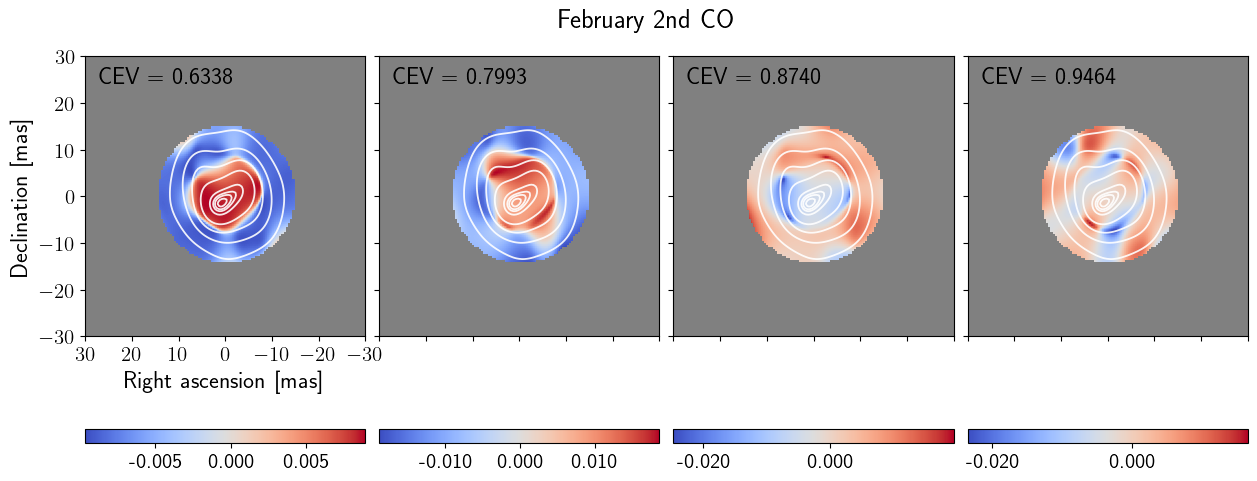}}
    \caption{First four Principal Components for the January and February epochs for the 1st CO (plots \textbf{a} and \textbf{b}), and the 2nd CO band heads (plots \textbf{e} and \textbf{f}). The relative scale of the structures in the eigen-images is displayed with a colorbar at the bottom of each panel. Only the central 20 mas of the eigen-images are shown on each panel. The white contours correspond to the mean image across wavelength (per given data set) and they represent 10, 30, 50, 70, 90, 95, 97, and 99$\%$ of the intensity's peak.}
    \label{fig:PCAC_conti}
\end{figure*}

\begin{figure*}
    \centering
    \includegraphics[width=16cm]{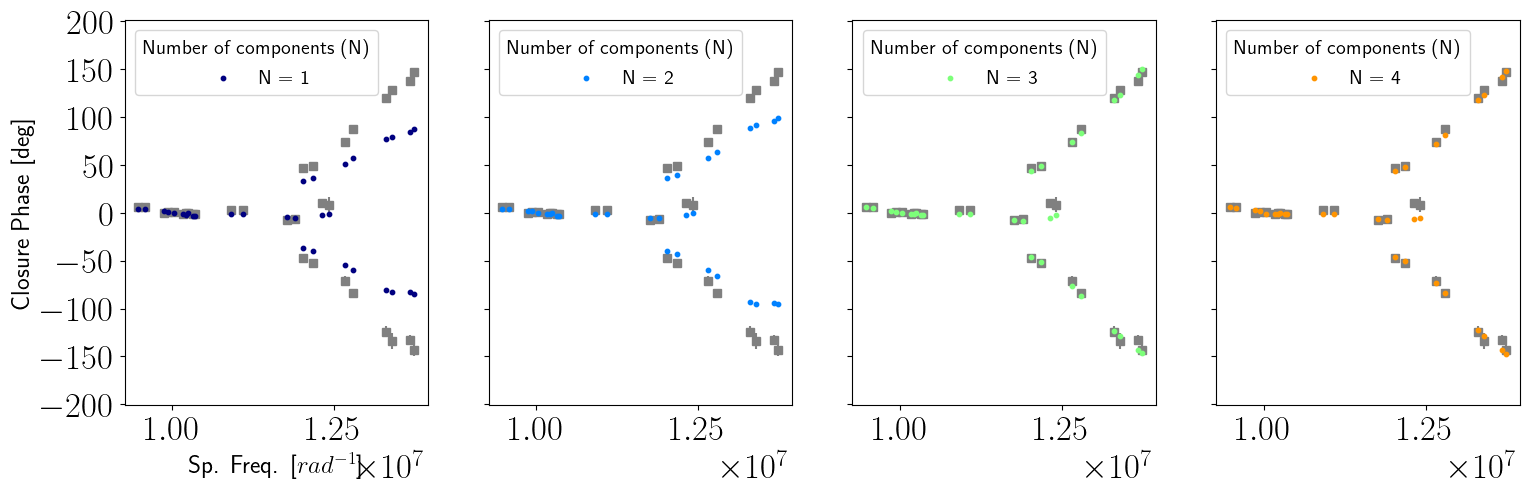}
    \caption{Representative example of the CPs versus spatial frequencies of the data set that corresponds to the 1st CO band-head at 2.2946 $\mu m$ (gray squares). The colored dots show the CPs from the recovered images obtained with different numbers of Principal Components (see label on the plot). }
    \label{fig:recovered_CPs}
\end{figure*}

The maps that correspond to the first four Principal Components of the CO band heads are reported in Fig. \ref{fig:PCAC_conti}. For both CO data-sets, we observe that the first two components trace the variance associated with the general changes in brightness and size observed in the images of the CO band heads across wavelengths. However, components 3-4 show considerably more asymmetric changes in the external structure of the target (for contours between 50\% and 10\% of the peak). The observed varying structures in components 3-4 are evenly distributed across the different position angles in the structure of the source. We suspect that the variance explained by these components is associated with the more asymmetric structure in the CO images. This could also be connected with the possibility of having a clumpy CO distribution of material in the envelope around R Car. As an example of the contribution of each component to the asymmetric structure of the target, in Figure \ref{fig:recovered_CPs}, we include the CPs obtained at a wavelength of 2.2946 $\mu m$ from the recovered images with cumulative components between 1-4. As it can be observed, the recovered images improve the recovery of the asymmetries traced by the CPs as we add more components to them. This is an indication that the variance observed in the eigen-images is tracing a more changing environment in the CO distribution across the sampled wavelengths. Unfortunately, the limited resolution of our observations is not enough to properly resolve these asymmetries completely. 

\subsection{Pure-line CO images}

\begin{figure}
    \centering
    \subfigure[]{\includegraphics[width=8cm]{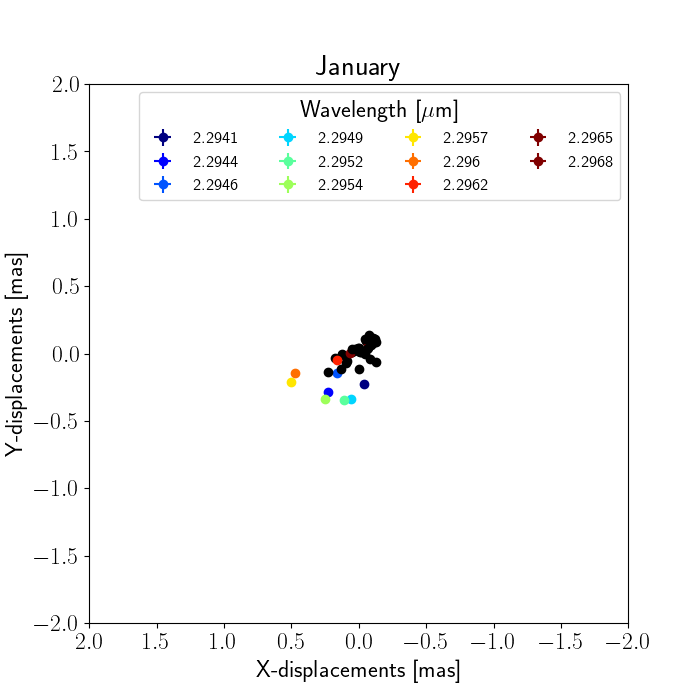}}
    \subfigure[]{\includegraphics[width=8cm]{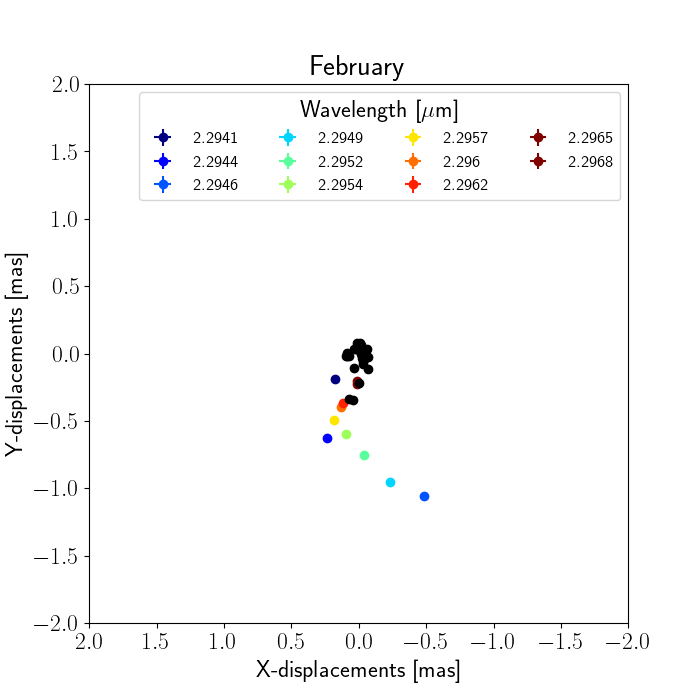}}
    \caption{2D photocenters offsets in mas. The black dots correspond to the photo-center position of the continuum emission and the labeled colored dots to the photo-centers of the CO band head. The error-bars of the reported displacements are smaller than the used symbols.}
    \label{fig:astro_offsets_1CO}
\end{figure}

 Our CO reconstructed images contain a contribution associated with the underlying stellar component. To remove such contribution, first, we center the images across the CO bandheads in a common reference position. For this purpose, we use the differential phases (DPs) in the GRAVITY data. DPs measure the photocenter displacement of an emission line with respect to the continuum for a given baseline length and orientation. We only restrict the pure-line CO images to the first band head, since the DPs of the second band head have much larger uncertainties. To compute the global 2-D photocenter displacement vector at a given spectral channel, we use the following relation \citep{kraus2012new}:

\begin{equation}\label{eq:p_vector}
    \overrightarrow{p} = -\frac{\phi_i}{2\pi}\cdot\frac{\lambda}{\overrightarrow{B_i}}\,,
\end{equation}

where $\phi_i$ is the GRAVITY differential phase measured on baseline $i$, $\overrightarrow{B_i}$ is the corresponding baseline vector [B$_x$, B$_y$] and $\lambda$ is the central wavelength of the working spectral channel. By solving the matrix represented by Eq. \ref{eq:p_vector}, we could find the global [p$_x$, p$_y$] displacements of the emission line.
We solve the system of equations employing the least-squares algorithm implemented in the \texttt{Python}  function \texttt{numpy.linalg.lstsq}. This function inverts the matrix formed by each one of the observed differential phases per baseline per channel to find the solution of the astrometric displacements. To obtain an estimation of the precision (i.e., 1$\sigma$ error-bars) of the reported displacements, we applied the same matrix inversion process to 100 different samples obtained with random points, using a Gaussian distribution with a standard deviation equivalent to the error-bars of the differential phases. The final astrometric displacements are equivalent to the mean of the 100 different samples, while their error-bars are equivalent to the standard deviation of the displacements from the samples.We obtained the displacements for the different spectral channels across the CO band heads, as well as for a few continuum channels. Fig. \ref{fig:astro_offsets_1CO} show the 1st CO band head offsets for our two epochs.
We can observe that the wavelengths corresponding to the continuum emission lie close to the reference position [0,0], and the ones of the band head show displacements within 1 mas. It is interesting to highlight that the CO centroid position of the two epochs is different, confirming the variability of the structure of the source between them. Once the centroid position of the images was computed, we use them to extract the pure-line CO images using the following method: 

\begin{itemize}
\item CPs are shift-invariant quantities, it means that images reconstructed with this observable could be placed at different positions in the pixel grid and, still, reproduce the CPs in the data. Therefore, to account for this effect, we computed the continuum and CO images, and we performed a sub-pixel shift to the corresponding centroid position obtained from the astrometric displacements [p$_x$, p$_y$] of the differential phases. 

\item For the subsequent analysis, we take the continuum images at $\lambda_{cont} = $ 2.228 $\mu m$. Once all the centroids of the images match with the astrometric displacements, we subtracted the continuum image from each one of the CO images across the different spectral channels. Since the total flux in the reconstructed images is normalized to the unity, before performing the subtraction, we calibrate the total flux on each image by scaling them with their corresponding normalized flux scale from the spectra in Figure \ref{fig:spectra_janfeb}. 
\end{itemize}

\begin{figure*}
    \centering
    \subfigure[]{\includegraphics[width=18.3cm]{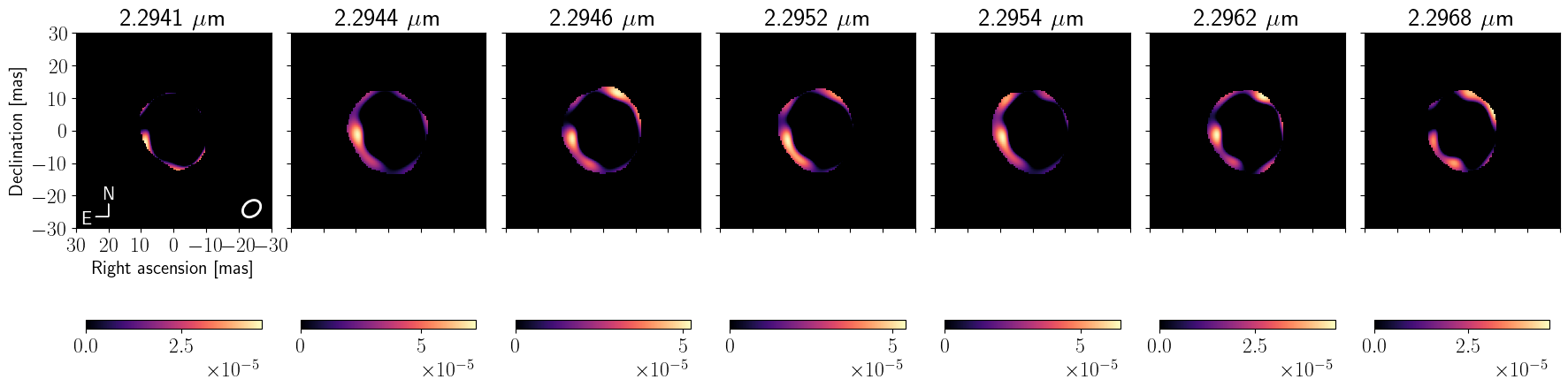}}
    \subfigure[]{ \includegraphics[width=18.3cm]{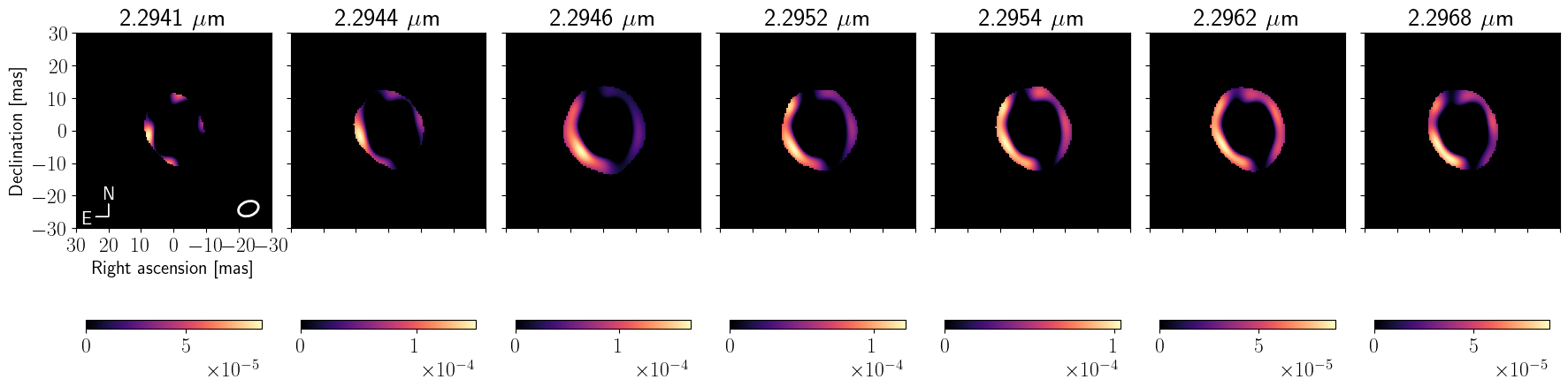}}
    \caption{ Pure-line CO images of the 1st CO band head for (a) the January and (b) the February epochs. The region that corresponds to the subtracted continuum contribution has been masked for a better visualization of the CO region. The white ellipse located in the lower right corner of the image at 2.2941 $\mu m$ corresponds to the mean synthesized primary beam. }
    \label{fig:pure_COimgs}
\end{figure*}

Figure \ref{fig:pure_COimgs} show the CO images with the continuum contribution subtracted. We masked the central region to better visualise the CO emitting region. From a visual inspection of the images, we can identify some features. First, there are different structures between wavelengths and epochs, suggesting the inhomogeneous nature of CO. The images also show bright and faint regions at different position angles. We relate them with zones of lower (faint regions) and higher (bright regions) densities and temperatures.

\section{Discussion}\label{sec:discussion}

 \subsection{CO column densities from the single-layer model}

The single-layer model described in Section \ref{sec:moslphere} allowed us to obtain information about the CO layer's physical conditions (e.g., opacity, temperature and sizes). Using the optical thickness and temperature derived for the layer, we estimate the corresponding column density, in units of $cm^{-2}$, by using the following relation  \citep{savage1991analysis}: 
\begin{equation}
N = \dfrac{\tau_L \mathrm{\Delta v} \sqrt{\pi}}{f_{ij}\lambda_{ij} c \dfrac{\pi e^2}{m_e c^2}},
\end{equation}
where $\tau_L$ is the optical thickness obtained from our single-layer model, $\mathrm{\Delta v}$ is the width of the band head in kms$^{-1}$, $f_{ij}$ is the absorption oscillator strength which depends on the temperature, $\lambda_{ij}$ is the wavelength of the transition, c is the speed of light, $e^-$ is the electron charge and m$_e$ its mass, the constant term $\pi$ e$^2$/m$_e$ c$^2$ = 8.8523$\times 10^{-13}$ cm/mol. To calculate the column density, we considered the following conditions:
\begin{itemize}
    \item We set $\lambda_{ij}$ = 2.2946 $\mu m$ to calculate the column density at the center of the first CO band head where the optical thickness has its largest value. 
    \item  Using Eq. (3) from \cite{goorvitch1994infrared}, we calculate f$_{ij}$ for the vibro-rotational transition V$_{0->2}$, J$_{20->21}$ (V and J are the vibrational and rotational states respectively). We obtained their corresponding Einstein coefficients, A, from \cite{chandra1996einstein}
    
    \item We calculated $\Delta v$ by fitting an inverse Lorentzian profile across the first CO band head for the two epochs. Since $\Delta v$ is expressed in kms$^{-1}$, $\Delta v$ = c$\Delta \lambda$/$\lambda_0$.  
    
\end{itemize}  

The obtained column densities for both epochs lie in the range $N_{\mathrm{CO}}$ $\sim$ 9.19$\times 10^{18}$ - 1.01$\times 10^{19}$ $\mathrm{cm}^{-2}$. This result has two implications. On one side, these values are consistent with the range of column densities reported by authors like \citet{rodriguez2021molsphere} and \citet{tsuji2006infrared} for other evolved stars. Furthermore, these values, on average, are high enough for shielding the dust particles allowing them to grow up to micron-size scales, where the stellar radiation could act efficiently to drive the observed winds in this kind of objects. On the other side, our current single-layer model returns large uncertainties in the estimated parameters. Therefore, we cannot confirm if CO formation happened between both epochs.

To obtain a direct comparison between the size of the photosphere ($\mathrm{R_*}$) and the size of the CO layer ($R_L$), we calculated the ratio $\mathrm{R_L/R_*}$. We take as reference the $\mathrm{R_*}$ value reported by \citet{monnier20142014} of 5 mas. From $\mathrm{R_L/R_*}$, we find a mean extension of the layer $\mathrm{R_L}$ $\sim$ 2.4 $\mathrm{R_*}$. To verify the consistency of our results, we compared them with other O-rich Miras for which the single-layer model has been applied. Table \ref{table:comparisons_otherM} shows the values of the layer temperature and $R_L/R*$ for the Mira stars R Leo, $\chi$ Cyg, and T Cep, as well as for our R Car estimates. From the values reported, we confirm that the parameters derived for R Car are consistent with other estimates reported in the literature for similar type of objects. 

\begin{table}[]
\centering
\caption{Parameters of the single-layer model for other AGBs}
\begin{threeparttable}
\begin{tabular}{cccc}
\hline\hline
Star                    & T$_*$ (K)                & T$_L$ (K)     & R$_L$/R$_*$                                            \\\hline
R Leo                   & 3856$\pm$119             & 1598$\pm$24   & 2.29$\pm$0.19 \\
$\chi$ Cyg                   & 3211$\pm$158             & 1737$\pm$53   & 1.91$\pm$0.02                             \\
T Cep                   & 3158$\pm$158             & 1685$\pm$53   & 2.15$\pm$0.02                             \\
R Car Jan (CO 2-0)  & 2800 & 1360 $^{+106}_{-90}$ & 2.43 $\pm$ 0.20                          \\[0.2cm]
R Car Feb (CO 2-0) &     2800                  & 1406 $^{+108}_{-83}$ & 2.42 $\pm$ 0.20                                                     \\[0.2cm]
R Car Jan (CO 3-1)  &         2800              & 1330 $^{+91}_{-81}$ & 2.44 $\pm$ 0.20                                                      \\[0.2cm]
R Car Feb (CO 3-1) &       2800                & 1401 $^{+122}_{-106}$ & 2.40
$\pm$ 0.20                                      \\[0.2cm] \hline
\end{tabular}
\begin{tablenotes}
\item {Parameters of the singe-layer model extracted from \cite{perrin2004unveiling} for two O-rich AGB stars (R Leo and T Cep) and one S-type ($\chi$ Cyg) compared with the parameters reported in this work for the CO band heads and GRAVITY epochs analyzed.}
\end{tablenotes}
\end{threeparttable}
\label{table:comparisons_otherM}
\end{table}

\subsection{Observational restrictions to the wind driving candidates
for M-type Mira stars}\label{subsec:slmodd}

Given the high chemical complexity around M-type AGB stars, it is not easy to determine the chemical composition of dust-driven winds. Even though observations tell us about the existence of different dust species in the circumstellar environment \citep{molster2010astromineralogy}, not all are efficient in absorbing/scattering the star's radiation to trigger a stellar wind. A condition for a dust grain to be considered a wind driver is that it must be formed closer to the stellar surface within reach of shock waves produced by the star. The closest distance from the star a dust grain can be formed depends on its chemical and optical properties \citep{Bladh_2012}.

As mentioned in Sec. \ref{sec:intro}, the observed CO vibro-rotational transitions (with $\Delta V$ = 2) originate in the deep photospheric layers where dust is forming. Therefore, with the single-layer results, we can set constrains on the types of dust grains that are considered wind-driving candidates and; which can coexist with similar physical conditions as the CO innermost layers detected with our GRAVITY observations. It is important to remember that for winds to be triggered, it is required that the star pulsations levitate the gas to a certain distance (defined as levitation distance, R$_\mathrm{l}$) where dust can condense (known as condensation distance, R$_\mathrm{c}$). \cite{Bladh_2012} define the levitation distance with the following expression: 

\begin{equation}\label{eq:r_levitation}
    \dfrac{R_\mathrm{l}}{R_\mathrm{*}} = \dfrac{R_\mathrm{0}}{R_\mathrm{*}}\left[1-\dfrac{R_\mathrm{0}}{R_\mathrm{*}}\left(\dfrac{u_\mathrm{0}}{u_\mathrm{esc}}\right)^{2}\right]^{-1}\,,
\end{equation}

where $u_\mathrm{0}$ is the initial velocity of the gas at a distance $R_\mathrm{0}$, and $u_\mathrm{esc}$ = (2GM$_\mathrm{*}$/R$_{*}$)$^{1/2}$ is the escape velocity. The levitation radius can be extracted directly from the size of the layer from our single-layer model or from the pure-line CO reconstructed images. The single-layer model provides us an average estimate of $R_\mathrm{l}$ = 2.43 R$_*$, since the model only considers a symmetric structure. However, the reconstructed images provide us a range of $R_\mathrm{l}$ between 2.25 and 2.72 R$_*$, due to the observed asymmetries. Assuming $u_\mathrm{esc}$ = 44 km s$^{-1}$, M = 0.87 M$_\odot$ \citep{takeuti2013method}, T$_\mathrm{eff}$ = 2800 K \citep{mcdonald2012fundamental}, R$_\mathrm{*}$ = 167 R$_\odot$ \citep{monnier20142014} for Eq.\,\ref{eq:r_levitation}; and that the CO band heads that we observed are formed in a region from the stellar surface at R$_\mathrm{0}$ = 2 R$_\mathrm{*}$, we computed an initial velocity of the CO of $u_\mathrm{0}$ = 13 km s$^{-1}$ considering the single-layer model and $u_\mathrm{0}$ between 10 and 16 km s$^{-1}$ using the CO reconstructed images. These values are consistent with the ones reported in the literature of the order of $u_\mathrm{0}\,\sim$  10 km s$^{-1}$ \citep{nowotny2010line}.

The dust condensation radius, R$_\mathrm{c}$, can be determined using the following expression:

\begin{equation}\label{eq:r_condensation}
    \dfrac{\mathrm{R_c}}{\mathrm{R_*}} = \dfrac{1}{2}\left(\dfrac{\mathrm{T_c}}{\mathrm{T_*}}\right)^{- \dfrac{4+\mathrm{p}}{2}} \quad \mathrm{where} \quad \kappa_\mathrm{abs} \propto \lambda^{-\mathrm{p}}\,,
\end{equation}

where $\mathrm{T_c}$ is the dust's condensation temperature, $\mathrm{T_*}$ is the star's effective temperature, and p is the dust absorption coefficient. The aforementioned equation considers a Planckian radiation field and that the dust grain temperature is determined by the condition of thermal equilibrium. For dust-driven winds to be triggered, gas must be levitated beyond the dust condensation radius. Therefore, we can obtain information for the type of dust grains that could be formed considering the R$_l$ values obtained from our single-layer model and the pure-line CO images. For this purpose, we estimated the condensation radii for several types of dust grains assuming the condensation coefficients and temperatures reported on \citet{Bladh_2012} and references therein. These values are reported in a p vs T$_c$ diagram. Considering that the levitation radius is smaller or equal to the condensation radius, we estimated a curve of possible p and T$_c$ values with the R$_l$ estimates from our single-layer model and pure-line CO reconstructed images. 

\begin{figure}
    \centering
    \includegraphics[width=7cm]{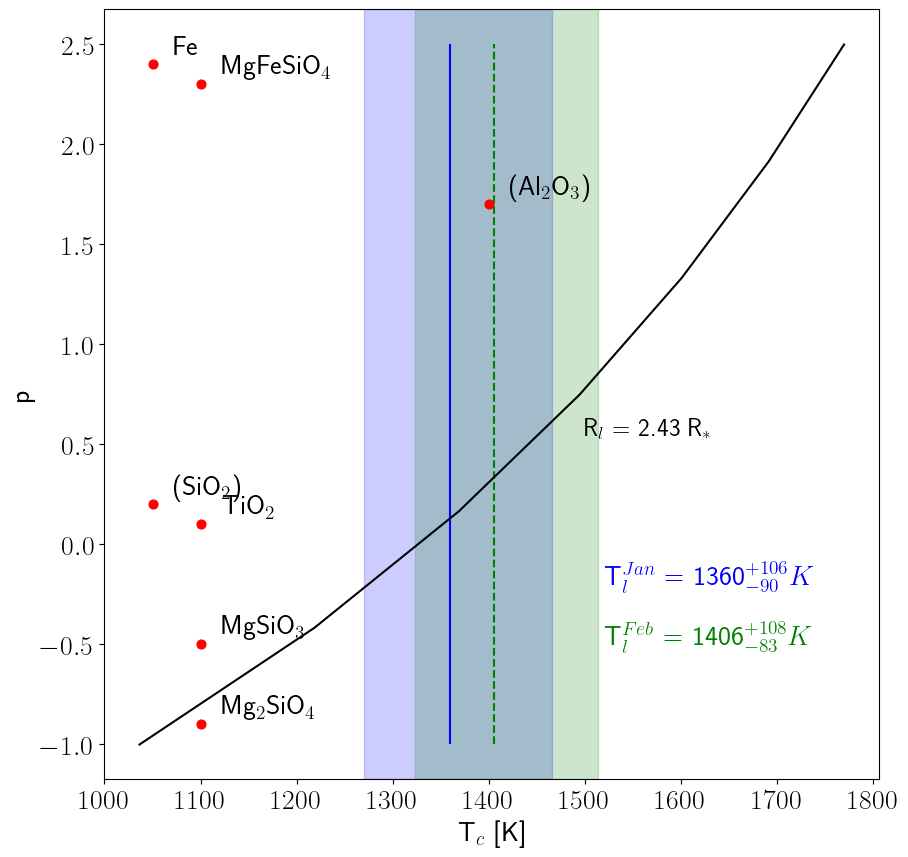}
    \caption{Curve of constant condensation radius (R$_c$ = 2.43 R$_*$) represented with a solid-black line. The plot shows the condensation temperature, T$_\mathrm{c}$, as a function of the dust absorption coefficient, p. The red circles indicate the positions of a variety of dust grains. The blue and green shaded regions correspond to the CO temperature ranges estimated from the single-layer model (see labels on the plot).}
    \label{fig:pTc_plane_RLayer}
\end{figure}

Figure \ref{fig:pTc_plane_RLayer} shows the T$_c$ vs p, diagram. The red dots correspond to a variety of dust grains with different condensation temperatures and values of p. The solid-black line displays the values of T$_c$ and p for the R$_l$ value obtained from the single-layer model. We observe that our CO layer radius is consistent with the condensation radius for composites Mg$_2$SiO$_4$ and MgSiO$_3$. However, we find layer temperatures around 1300 $\pm$ 100 K (blue and green shaded regions on Fig.\ref{fig:pTc_plane_RLayer}), which are slightly higher than the expected condensation temperature for those two dust composites (T$_c\,\sim$ 1100 K). This temperature discrepancy could be explained by the fact that the single-layer model considers a homogeneous and symmetric distribution of the CO layer. However, the reconstructed images clearly show that the CO distribution is neither homogeneous nor symmetric. Therefore, regions with lower temperatures and larger levitation radius could exist around R Car.

\begin{figure}
    \centering
    \subfigure[]{\includegraphics[width=7cm]{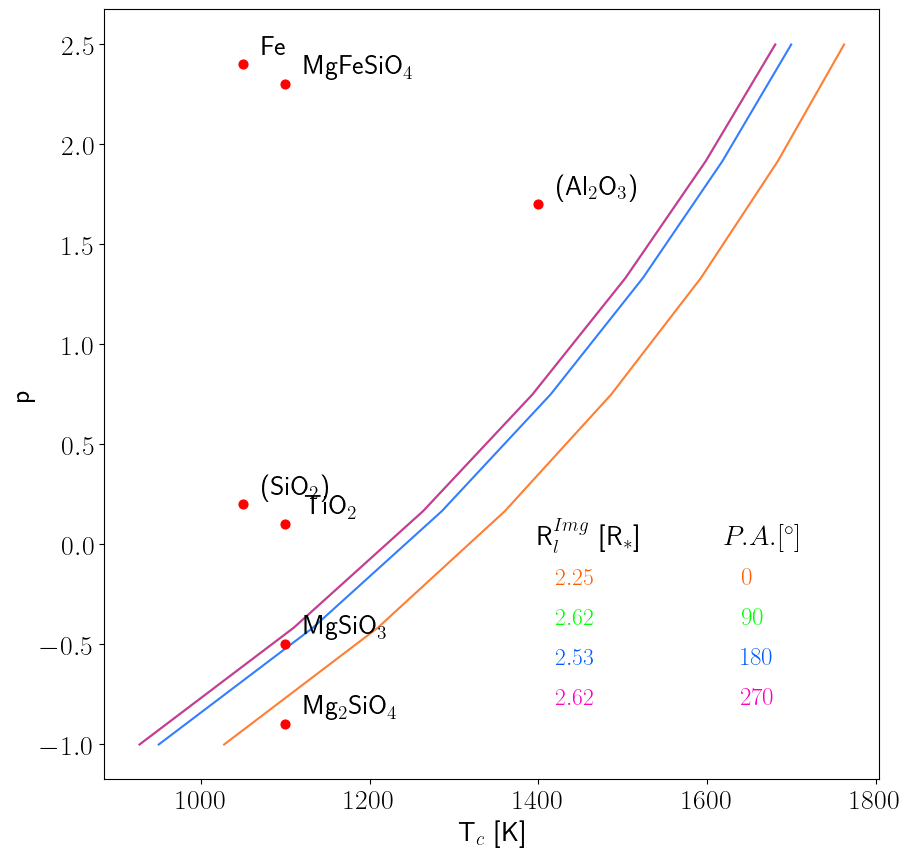}}
    \subfigure[]{\includegraphics[width=7cm]{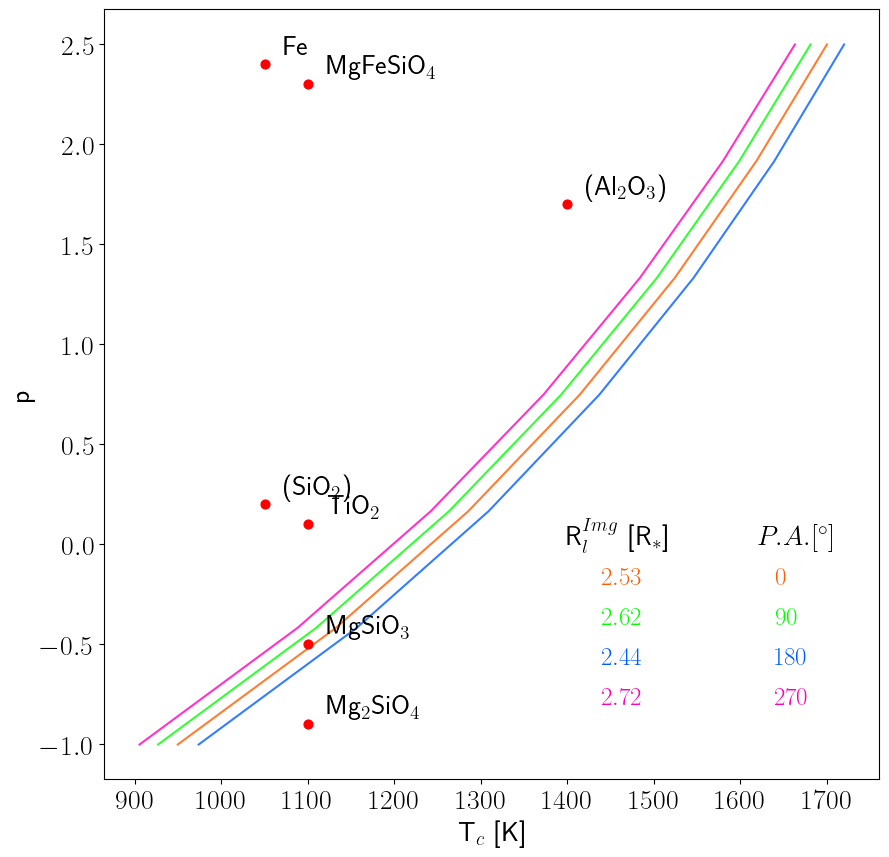}}
    \caption{The panels are similar to Fig. \ref{fig:pTc_plane_RLayer} but they show the the median condensation radii obtained directly from the pure-line CO reconstructed images at different position angles (see labels on the panels) for a) the January and b) February epochs.}
    \label{fig:pTc_images}
\end{figure}

Figure \ref{fig:pTc_images} displays the T$_c$ vs p diagrams using the R$_l$ values extracted from the pure-line CO images for both observed epochs. The diagrams show the CO levitation radii at different orientations: 0$^{\circ}$, 90$^{\circ}$, 180$^{\circ}$, and 270$^{\circ}$. Since we have seven images per epoch, we adopted the median value per image data set at each orientation. Similarly to Fig. \ref{fig:pTc_plane_RLayer},  Mg$_2$SiO$_4$ and MgSiO$_3$ dust composites are consistent with the layer's radii. However, we notice that depending on the position angle, either one or both of them lie below the estimated condensation/levitation radius. This supports the hypothesis that dust formation could occur in an anisotropic way over the CO shell extension. 

Magnesium-iron silicates (olivines and pyroxenes) with nano-grain sizes ($\sim$ 0.1 $\mu$as) have been detected already in the circumstellar environments of M-type AGB stars \citep{norris2012close, ohnaka_2017}, and they appear to be the main responsible for the dust-driven winds observed in these stars. While our GRAVITY observations and models suggest the possible coexistence of those dust types at the inner gaseous (CO) shells , we require additional interferometric observations in the mid-infrared (potentially with the  beam-combiner MATISSE-VLTI) to confirm this scenario. Very recent models of dusty winds in M-type AGBs suggest Fe enrichment of the wind as it moves far away from the photosphere \citep{hofner_2022}. Those models suggest an inner core of Mg$_2$SiO$_4$ (or of a similar composite) surrounded by more outer layers of MgFeSiO$_4$. New MATISSE-VLTI interferometric observations could help us to confirm this scenario in the proximity of R Car in the near-future. By confirming this scenario, mid-infrared interferometric observations could shed new light into the problem of a lack of the 10 $\mu$m silicate feature in the spectra of M-type AGB stars.

\subsection{The role of CO asymmetries on mass-loss estimates}

The mass-loss rate is a crucial parameter for understanding the evolution of M-type AGB stars. \citet{bladh2019extensive} present an extensive grid of DARWIN dynamic atmospheres-and-wind models to establish the mass-loss rates, wind velocities and grain sizes for M-type AGB stars. These models assume that the principal mechanism for wind triggering is the radiation scattering on Mg$_2$SiO$_4$ composites. This grid of models produce mass-loss rates of the order of 1$\times$10$^{-6}$ M$_{\odot}$ yr$^{-1}$ for stellar parameters similar to R Car (e.g., M$_*$ = 0.75 M$_{\odot}$; log[L$_*$/L$_{\odot}$] = 3.55; P = 283 days). 

More importantly, those models show that, for the range of mass (0.75 - 1 M$_{\odot}$) and luminosity (log[L$_*$/L$_{\odot}$] = 3.7) of R Car, large values of the pulsation amplitude in the models and high chemical enrichment develop stable winds based on the Mg$_2$SiO$_4$ composites. However, there is a region (when $\Delta u_p$ = 2 km/s and n$_d$/n$_H$ = 1$\times$10$^{15}$) of low chemical enrichment and small pulsation amplitudes where those DARWIN models do not develop stable dust-driven winds. While those models are important to demonstrate the dynamical evolution of dust-driven winds with chemical properties based on magnesium-iron silicates, they do not incorporate the importance of asymmetries on the development of the wind. 

\citet{liljegren2018atmospheres} studied the effects of complex convective structures on wind production and mass loss in M-type AGB stars. The authors use the CO5BOLD code \citep{freytag2003co5bold} to simulate the convective surface and pulsations of AGBs, using the results as boundary conditions in their DARWIN models. Those models show that gas is not levitated uniformly above the star's surface (only about 70$\%$ or less of the stellar surface is covered), causing dust not to be formed in a spherical layer around the star. Instead, it forms clumps. This appears to have a direct implication in an overestimation of the mass-loss rate determined with homogeneous models. In this context, our work shows direct evidence of the asymmetric and variable gas distribution at the innermost layers of M-type AGB stars. A specific DARWIN model of R Car is beyond the scope of this paper and it is left for future studies. However, by contrasting the structures observed in our reconstructed images and future dedicated CO5BOLD/DARWIN models we could obtain better constraints on the mass-loss evolution and the role of Magnesium dust composites on the development of stable winds. 

\section{Conclusions}\label{sec:conclusions}

In this work, we reported an analysis of the GRAVITY-VLTI ($K-$band) interferometric data of  the M-type Mira star, R Car. Our primary scientific goal is to provide new observational evidence on the mass-loss mechanisms associated with M-type AGB Mira stars. In particular, we were interested in linking the structure of the gas around R Car with the origin of the dusty winds observed in M-type Mira stars. For this purpose, we characterize the morphology of the target within $\sim$3 stellar radii. 

To obtain a first-order estimate of the size of the source, we applied a geometrical model to the V$^2$ across the K-band for our two epochs of observation with GRAVITY-VLTI, which are separated by $\sim$ 30 days that correspond to $\sim$ 10\% of the full pulsation period. We find that the size of the target changes with wavelength due to the presence of molecules like H$_2$O and CO at $\sim 2\, \mu m$ and $\sim 2.29-2.33\, \mu m$, respectively. The best estimate of the stellar disk diameter in $K-$band was derived at a wavelength $\sim 2.23\, \mu m$. Our parametric model allows us to obtain a diameter of 16.67$\pm 0.05$ mas (3.03 au) in January, 2018 and, 14.84$\pm 0.06$ mas (2.70 au) in February, 2018. 

We used a parametric single-layer model, which allowed us to have a physical estimate of the size, temperature and optical thickness of the CO layer.  Using the estimated optical thickness and temperature at the minimum peak of the first CO band head, we derived a column density N$_\mathrm{CO}$ $\sim$ 9$\times$10$^{18}$ - 1$\times$10$^{19}$cm$^{-2}$. This estimate is in agreement with the column densities reported for other evolved stars and it is high enough for allowing coalescence dust formation. The precision of our geometrical single-layer model is not enough to confirm the formation of CO. For this, higher angular resolution measurements in combination with more complex models of the CO layer would allow us to better constrain the sizes, temperatures and optical thicknesses of the gas, which would result in better estimations of the column densities.

From the temperature and size of the CO layer, we set constraints on the dust candidates that could coexist with the gas. We find that the composites Mg$_2$SiO$_3$ and MgSiO$_4$ could coexist at the same distances as the ones we obtained for the CO innermost layers ($\sim$ 2 - 2.4 R$_*$). However, the condensation temperatures for these dust particles are lower than the temperatures obtained from the CO layer model. We propose that the temperature discrepancy is due to the nature of the single-layer model, which considers a homogeneous and symmetric distribution of the CO layer, giving mean values for T$_\mathrm{L}$, R$_\mathrm{L}$ and $\tau_\mathrm{L}$. Our hypothesis is supported by the varying interferometric CPs and the posterior image reconstruction, which show that the CO distribution is neither homogeneous nor symmetric.

To complement the information obtained from our symmetric parametric models, we reconstruct interferometric images from the GRAVITY data.  As we expected from the variation of the CPs, we observed several asymmetries in the images across the spectral ranges and for both epochs of observation. The presence of those asymmetries were confirmed (above 3 $\sigma$ detection level) using bootstrap, PCA and spectro-astrometric analyses. We propose that these asymmetries trace a complex, perhaps clumpy structure of the CO molecule distribution, which  also changes within  a period of 30 days or less. 

Recent models of the convection and pulsation mechanisms of M-type AGB stars support that an asymmetric levitation of the gas at the innermost regions of these stars has a direct impact on the estimation of mass-loss rates and the formation of stable dust-driven winds. A dedicated characterization of the asymmetries observed in the CO distribution of R Car, together with new GRAVITY-VLTI data (with the longest baselines available at the VLTI) could provide unique observational evidence to test this scenario. Furthermore, complementary mid-infrared interferometric data with MATISSE-VLTI could help us to test the role of Magnesium based dust composites in the formation of dust-driven winds in M-type Mira stars and their subsequent chemical evolution.



\begin{acknowledgements}
The authors of this work thank the anonymous referee for his/her valuable comments to improve our manuscript. A.R.-G. acknowledges the support received through the PhD scholarship (No. 760678) granted by the Mexican Council of Science CONACyT. J.S.-B. acknowledges the support received from the UNAM PAPIIT projects IA 101220 and IA 105023; and from the CONACyT “Ciencia de Frontera” project CF-2019/263975. R.S. acknowledges financial support from the State Agency for Research of the Spanish MCIU through the “Center of Excellence Severo Ochoa” award for the Instituto de Astrofísica de Andalucía (SEV-2017-0709). R.S. acknowledges financial support from national project PGC2018-095049-B-C21 (MCIU/AEI/FEDER, UE). This project is based on observations collected at the European Southern Observatory under ESO programme 0100.D-0835.
      
\end{acknowledgements}

%
%
\bibliographystyle{aa}
\bibliography{bibliog.bib}

\begin{thebibliography}{64}
\expandafter\ifx\csname natexlab\endcsname\relax\def\natexlab#1{#1}\fi

\bibitem[{Abuter {et~al.}(2017)Abuter, Accardo, Amorim, Anugu, Avila, Azouaoui,
  Benisty, Berger, Blind, Bonnet, {et~al.}}]{abuter2017first}
Abuter, R., Accardo, M., Amorim, A., {et~al.} 2017, Astronomy \& Astrophysics,
  602, A94

\bibitem[{Allard {et~al.}(2010)Allard, Homeier, \& Freytag}]{allard2010model}
Allard, F., Homeier, D., \& Freytag, B. 2010, arXiv preprint arXiv:1011.5405

\bibitem[{Babu \& Singh(1983)}]{babu1983inference}
Babu, G.~J. \& Singh, K. 1983, The Annals of Statistics, 999

\bibitem[{Baron \& Young(2008)}]{baron2008image}
Baron, F. \& Young, J.~S. 2008, in Optical and Infrared Interferometry, Vol.
  7013, International Society for Optics and Photonics, 70133X

\bibitem[{Berger \& Segransan(2007)}]{berger2007introduction}
Berger, J.~P. \& Segransan, D. 2007, New Astronomy Reviews, 51, 576

\bibitem[{{Bladh} \& {H{\"o}fner}(2012)}]{Bladh_2012}
{Bladh}, S. \& {H{\"o}fner}, S. 2012, \aap, 546, A76

\bibitem[{Bladh {et~al.}(2019)Bladh, Liljegren, H{\"o}fner, Aringer, \&
  Marigo}]{bladh2019extensive}
Bladh, S., Liljegren, S., H{\"o}fner, S., Aringer, B., \& Marigo, P. 2019,
  Astronomy \& Astrophysics, 626, A100

\bibitem[{Chandra {et~al.}(1996)Chandra, Maheshwari, \&
  Sharma}]{chandra1996einstein}
Chandra, S., Maheshwari, V., \& Sharma, A. 1996, Astronomy and Astrophysics
  Supplement Series, 117, 557

\bibitem[{Cruzal{\`e}bes {et~al.}(2015)Cruzal{\`e}bes, Jorissen, Chiavassa,
  Paladini, Rabbia, \& Spang}]{cruzalebes2015departure}
Cruzal{\`e}bes, P., Jorissen, A., Chiavassa, A., {et~al.} 2015, Monthly Notices
  of the Royal Astronomical Society, 446, 3277

\bibitem[{Foreman-Mackey {et~al.}(2013)Foreman-Mackey, Hogg, Lang, \&
  Goodman}]{foreman2013emcee}
Foreman-Mackey, D., Hogg, D.~W., Lang, D., \& Goodman, J. 2013, Publications of
  the Astronomical Society of the Pacific, 125, 306

\bibitem[{Freytag {et~al.}(2002)Freytag, Steffen, \& Dorch}]{freytag2002spots}
Freytag, B., Steffen, M., \& Dorch, B. 2002, Astronomische Nachrichten, 323,
  213

\bibitem[{Freytag {et~al.}(2003)Freytag, Steffen, Wedemeyer-B{\"o}hm, Ludwig,
  Leenaarts, \& Schaffenberger}]{freytag2003co5bold}
Freytag, B., Steffen, M., Wedemeyer-B{\"o}hm, S., {et~al.} 2003, CO5BOLD User
  Manual

\bibitem[{{Gaia Collaboration}(2020)}]{2020yCat.1350....0G}
{Gaia Collaboration}. 2020, VizieR Online Data Catalog, I/350

\bibitem[{Gautschy-Loidl {et~al.}(2004)Gautschy-Loidl, H{\"o}fner,
  J{\o}rgensen, \& Hron}]{gautschy2004dynamic}
Gautschy-Loidl, R., H{\"o}fner, S., J{\o}rgensen, U.~G., \& Hron, J. 2004,
  Astronomy \& Astrophysics, 422, 289

\bibitem[{Geballe {et~al.}(2007)Geballe, Evans, van Loon, Smalley, Rushton, \&
  Eyres}]{geballe2007infrared}
Geballe, T., Evans, A., van Loon, J.~T., {et~al.} 2007, in The Nature of V838
  Mon and its Light Echo, Vol. 363, 110

\bibitem[{Goorvitch(1994)}]{goorvitch1994infrared}
Goorvitch, D. 1994, The Astrophysical Journal Supplement Series, 95, 535

\bibitem[{Gull \& Skilling(1984)}]{gull1984max}
Gull, S.~F. \& Skilling, J. 1984, in IEE Proceedings F-Communications, Radar
  and Signal Processing, Vol. 131, IET, 646--659

\bibitem[{Haubois {et~al.}(2015)Haubois, Wittkowski, Perrin, Kervella,
  M{\'e}rand, Thi{\'e}baut, Ridgway, Ireland, \& Scholz}]{haubois2015resolving}
Haubois, X., Wittkowski, M., Perrin, G., {et~al.} 2015, Astronomy \&
  Astrophysics, 582, A71

\bibitem[{H{\"o}fner(2008)}]{hofner2008winds}
H{\"o}fner, S. 2008, Astronomy \& Astrophysics, 491, L1

\bibitem[{{H{\"o}fner} {et~al.}(2022){H{\"o}fner}, {Bladh}, {Aringer}, \&
  {Eriksson}}]{hofner_2022}
{H{\"o}fner}, S., {Bladh}, S., {Aringer}, B., \& {Eriksson}, K. 2022, \aap,
  657, A109

\bibitem[{H{\"o}fner \& Olofsson(2018)}]{hofner2018mass}
H{\"o}fner, S. \& Olofsson, H. 2018, The Astronomy and Astrophysics Review, 26,
  1

\bibitem[{Ireland {et~al.}(2004)Ireland, Tuthill, Bedding, Robertson, \&
  Jacob}]{ireland2004multiwavelength}
Ireland, M., Tuthill, P., Bedding, T., Robertson, J., \& Jacob, A. 2004,
  Monthly Notices of the Royal Astronomical Society, 350, 365

\bibitem[{Ireland {et~al.}(2008)Ireland, Scholz, \&
  Wood}]{ireland2008dynamical}
Ireland, M.~J., Scholz, M., \& Wood, P.~R. 2008, Monthly Notices of the Royal
  Astronomical Society, 391, 1994

\bibitem[{Ireland {et~al.}(2011)Ireland, Scholz, \&
  Wood}]{ireland2011dynamical}
Ireland, M.~J., Scholz, M., \& Wood, P.~R. 2011, Monthly Notices of the Royal
  Astronomical Society, 418, 114

\bibitem[{{Jeong} {et~al.}(2003){Jeong}, {Winters}, {Le Bertre}, \&
  {Sedlmayr}}]{Jeong_2003}
{Jeong}, K.~S., {Winters}, J.~M., {Le Bertre}, T., \& {Sedlmayr}, E. 2003,
  \aap, 407, 191

\bibitem[{Kraus(2012)}]{kraus2012new}
Kraus, S. 2012, in Optical and Infrared Interferometry III, Vol. 8445,
  International Society for Optics and Photonics, 84451H

\bibitem[{Lacour {et~al.}(2009)Lacour, Thi{\'e}baut, Perrin, Meimon, Haubois,
  Pedretti, Ridgway, Monnier, Berger, Schuller, {et~al.}}]{lacour2009pulsation}
Lacour, S., Thi{\'e}baut, E., Perrin, G., {et~al.} 2009, The Astrophysical
  Journal, 707, 632

\bibitem[{Lapeyrere {et~al.}(2014)Lapeyrere, Kervella, Lacour, Azouaoui,
  Garcia-Dabo, Perrin, Eisenhauer, Perraut, Straubmeier, Amorim,
  {et~al.}}]{lapeyrere2014gravity}
Lapeyrere, V., Kervella, P., Lacour, S., {et~al.} 2014, in Optical and Infrared
  Interferometry IV, Vol. 9146, International Society for Optics and Photonics,
  91462D

\bibitem[{Le~Bouquin {et~al.}(2011)Le~Bouquin, Berger, Lazareff, Zins,
  Haguenauer, Jocou, Kern, Millan-Gabet, Traub, Absil,
  {et~al.}}]{le2011pionier}
Le~Bouquin, J.-B., Berger, J.-P., Lazareff, B., {et~al.} 2011, Astronomy \&
  Astrophysics, 535, A67

\bibitem[{Liljegren {et~al.}(2018)Liljegren, H{\"o}fner, Freytag, \&
  Bladh}]{liljegren2018atmospheres}
Liljegren, S., H{\"o}fner, S., Freytag, B., \& Bladh, S. 2018, Astronomy \&
  Astrophysics, 619, A47

\bibitem[{Liljegren {et~al.}(2016)Liljegren, H{\"o}fner, Nowotny, \&
  Eriksson}]{liljegren2016dust}
Liljegren, S., H{\"o}fner, S., Nowotny, W., \& Eriksson, K. 2016, Astronomy \&
  Astrophysics, 589, A130

\bibitem[{McDonald {et~al.}(2012)McDonald, Zijlstra, \&
  Boyer}]{mcdonald2012fundamental}
McDonald, I., Zijlstra, A.~A., \& Boyer, M.~L. 2012, Monthly Notices of the
  Royal Astronomical Society, 427, 343

\bibitem[{Medeiros {et~al.}(2018)Medeiros, Lauer, Psaltis, \&
  {\"O}zel}]{medeiros2018principal}
Medeiros, L., Lauer, T.~R., Psaltis, D., \& {\"O}zel, F. 2018, The
  Astrophysical Journal, 864, 7

\bibitem[{Molster {et~al.}(2010)Molster, Waters, \&
  Kemper}]{molster2010astromineralogy}
Molster, F., Waters, L., \& Kemper, F. 2010, T. Henning (Lecture Notes in
  Physics, Vol. 815

\bibitem[{Monnier {et~al.}(2014)Monnier, Berger, Le~Bouquin, Tuthill,
  Wittkowski, Grellmann, M{\"u}ller, Renganswany, Hummel, Hofmann,
  {et~al.}}]{monnier20142014}
Monnier, J.~D., Berger, J.-P., Le~Bouquin, J.-B., {et~al.} 2014, in Optical and
  Infrared Interferometry IV, Vol. 9146, International Society for Optics and
  Photonics, 91461Q

\bibitem[{Montarg{\`e}s {et~al.}(2014)Montarg{\`e}s, Kervella, Perrin, Ohnaka,
  Chiavassa, Ridgway, \& Lacour}]{montarges2014properties}
Montarg{\`e}s, M., Kervella, P., Perrin, G., {et~al.} 2014, Astronomy \&
  Astrophysics, 572, A17

\bibitem[{Norris {et~al.}(2012)Norris, Tuthill, Ireland, Lacour, Zijlstra,
  Lykou, Evans, Stewart, \& Bedding}]{norris2012close}
Norris, B.~R., Tuthill, P.~G., Ireland, M.~J., {et~al.} 2012, Nature, 484, 220

\bibitem[{Nowotny {et~al.}(2005)Nowotny, Aringer, H{\"o}fner, Gautschy-Loidl,
  \& Windsteig}]{nowotny2005atmospheric}
Nowotny, W., Aringer, B., H{\"o}fner, S., Gautschy-Loidl, R., \& Windsteig, W.
  2005, Astronomy \& Astrophysics, 437, 273

\bibitem[{Nowotny {et~al.}(2010)Nowotny, H{\"o}fner, \&
  Aringer}]{nowotny2010line}
Nowotny, W., H{\"o}fner, S., \& Aringer, B. 2010, Astronomy \& Astrophysics,
  514, A35

\bibitem[{{Ohnaka} {et~al.}(2017){Ohnaka}, {Weigelt}, \&
  {Hofmann}}]{ohnaka_2017}
{Ohnaka}, K., {Weigelt}, G., \& {Hofmann}, K.~H. 2017, \aap, 597, A20

\bibitem[{Paladini(2011)}]{paladini2011interferometry}
Paladini, C. 2011, Interferometry of carbon rich AGB stars (na)

\bibitem[{Paladini {et~al.}(2018)Paladini, Baron, Jorissen, Le~Bouquin,
  Freytag, Van~Eck, Wittkowski, Hron, Chiavassa, Berger,
  {et~al.}}]{paladini2018large}
Paladini, C., Baron, F., Jorissen, A., {et~al.} 2018, Nature, 553, 310

\bibitem[{Pedregosa {et~al.}(2011)Pedregosa, Varoquaux, Gramfort, Michel,
  Thirion, Grisel, Blondel, Prettenhofer, Weiss, Dubourg,
  {et~al.}}]{pedregosa2011scikit}
Pedregosa, F., Varoquaux, G., Gramfort, A., {et~al.} 2011, the Journal of
  machine Learning research, 12, 2825

\bibitem[{Perrin {et~al.}(2020)Perrin, Ridgway, Lacour, Haubois, Thi{\'e}baut,
  Berger, Lacasse, Millan-Gabet, Monnier, Pedretti,
  {et~al.}}]{perrin2020evidence}
Perrin, G., Ridgway, S., Lacour, S., {et~al.} 2020, Astronomy \& Astrophysics,
  642, A82

\bibitem[{Perrin {et~al.}(2004)Perrin, Ridgway, Mennesson, Cotton, Woillez,
  Verhoelst, Schuller, Traub, Millan-Gabet, \& Lacasse}]{perrin2004unveiling}
Perrin, G., Ridgway, S., Mennesson, B., {et~al.} 2004, Astronomy \&
  Astrophysics, 426, 279

\bibitem[{Perrin {et~al.}(2005)Perrin, Ridgway, Verhoelst, Schuller, Traub,
  Millan-Gabet, \& Lacasse}]{perrin2005study}
Perrin, G., Ridgway, S., Verhoelst, T., {et~al.} 2005, Astronomy \&
  Astrophysics, 436, 317

\bibitem[{Ragland {et~al.}(2006)Ragland, Traub, Berger, Danchi, Monnier,
  Willson, Carleton, Lacasse, Millan-Gabet, Pedretti,
  {et~al.}}]{ragland2006first}
Ragland, S., Traub, W., Berger, J.-P., {et~al.} 2006, The Astrophysical
  Journal, 652, 650

\bibitem[{Rau {et~al.}(2017)Rau, Hron, Paladini, Aringer, Eriksson, Marigo,
  Nowotny, \& Grellmann}]{rau2017adventure}
Rau, G., Hron, J., Paladini, C., {et~al.} 2017, Astronomy \& astrophysics, 600,
  A92

\bibitem[{Rau {et~al.}(2015)Rau, Paladini, Hron, Aringer, Groenewegen, \&
  Nowotny}]{rau2015modelling}
Rau, G., Paladini, C., Hron, J., {et~al.} 2015, Astronomy \& astrophysics, 583,
  A106

\bibitem[{Renard {et~al.}(2011)Renard, Thi{\'e}baut, \&
  Malbet}]{renard2011image}
Renard, S., Thi{\'e}baut, E., \& Malbet, F. 2011, Astronomy \& Astrophysics,
  533, A64

\bibitem[{Rodr{\'\i}guez-Coira {et~al.}(2021)Rodr{\'\i}guez-Coira, Paumard,
  Perrin, Vincent, Abuter, Amorim, Baub{\"o}ck, Berger, Bonnet, Brandner,
  {et~al.}}]{rodriguez2021molsphere}
Rodr{\'\i}guez-Coira, G., Paumard, T., Perrin, G., {et~al.} 2021, arXiv
  preprint arXiv:2105.09832

\bibitem[{Sacuto {et~al.}(2011)Sacuto, Aringer, Hron, Nowotny, Paladini,
  Verhoelst, \& H{\"o}fner}]{sacuto2011observing}
Sacuto, S., Aringer, B., Hron, J., {et~al.} 2011, Astronomy \& Astrophysics,
  525, A42

\bibitem[{Sanchez-Bermudez {et~al.}(2018)Sanchez-Bermudez, Millour, Baron, van
  Boekel, Bourg{\`e}s, Duvert, Garcia, Gomes, Hofmann, Henning,
  {et~al.}}]{sanchez2018chromatic}
Sanchez-Bermudez, J., Millour, F., Baron, F., {et~al.} 2018, Experimental
  Astronomy, 46, 457

\bibitem[{Savage \& Sembach(1991)}]{savage1991analysis}
Savage, B.~D. \& Sembach, K.~R. 1991, The Astrophysical Journal, 379, 245

\bibitem[{Takeuti {et~al.}(2013)Takeuti, Nakagawa, Kurayama, \&
  Honma}]{takeuti2013method}
Takeuti, M., Nakagawa, A., Kurayama, T., \& Honma, M. 2013, Publications of the
  Astronomical Society of Japan, 65, 60

\bibitem[{Tsuji(2006)}]{tsuji2006infrared}
Tsuji, T. 2006, The Astrophysical Journal, 645, 1448

\bibitem[{{Wittkowski} {et~al.}(2008){Wittkowski}, {Boboltz}, {Driebe}, {Le
  Bouquin}, {Millour}, {Ohnaka}, \& {Scholz}}]{Wittkowski_2008}
{Wittkowski}, M., {Boboltz}, D.~A., {Driebe}, T., {et~al.} 2008, \aap, 479, L21

\bibitem[{{Wittkowski} {et~al.}(2011){Wittkowski}, {Boboltz}, {Ireland},
  {Karovicova}, {Ohnaka}, {Scholz}, {van Wyk}, {Whitelock}, {Wood}, \&
  {Zijlstra}}]{Wittkowski_2011}
{Wittkowski}, M., {Boboltz}, D.~A., {Ireland}, M., {et~al.} 2011, \aap, 532, L7

\bibitem[{{Wittkowski} {et~al.}(2016){Wittkowski}, {Chiavassa}, {Freytag},
  {Scholz}, {H{\"o}fner}, {Karovicova}, \& {Whitelock}}]{Wittkowski_2016}
{Wittkowski}, M., {Chiavassa}, A., {Freytag}, B., {et~al.} 2016, \aap, 587, A12

\bibitem[{Wittkowski {et~al.}(2012)Wittkowski, Hauschildt, Arroyo-Torres, \&
  Marcaide}]{wittkowski2012fundamental}
Wittkowski, M., Hauschildt, P., Arroyo-Torres, B., \& Marcaide, J. 2012,
  Astronomy \& Astrophysics, 540, L12

\bibitem[{Wittkowski {et~al.}(2017)Wittkowski, Hofmann, H{\"o}fner, Le~Bouquin,
  Nowotny, Paladini, Young, Berger, Brunner, de~Gregorio-Monsalvo,
  {et~al.}}]{wittkowski2017aperture}
Wittkowski, M., Hofmann, K.-H., H{\"o}fner, S., {et~al.} 2017, Astronomy \&
  astrophysics, 601, A3

\bibitem[{Wittkowski \& Paladini(2014)}]{wittkowski2014atmosphere}
Wittkowski, M. \& Paladini, C. 2014, European Astronomical Society Publications
  Series, 70, 179

\bibitem[{Wittkowski {et~al.}(2018)Wittkowski, Rau, Chiavassa, H{\"o}fner,
  Scholz, Wood, de~Wit, Eisenhauer, Haubois, \& Paumard}]{wittkowski2018vlti}
Wittkowski, M., Rau, G., Chiavassa, A., {et~al.} 2018, Astronomy \&
  Astrophysics, 613, L7

\bibitem[{Zoubir \& Iskander(2004)}]{zoubir2004bootstrap}
Zoubir, A.~M. \& Iskander, D.~R. 2004, Bootstrap techniques for signal
  processing (Cambridge University Press)

\end{thebibliography}

\begin{appendix}
\section{Gravity observations' list and observables }{\label{sec:observations_grav}}

\begin{table*}[]
\centering
\caption{Observation log of our GRAVITY R Car observations. HD80404 was selected as a calibrator star. The coherence time ($\tau_0$) was taken from the Paranal Astronomical Site Monitoring (ASM)} 
\begin{tabular}{llllllll}
No. & Date & \begin{tabular}[c]{@{}l@{}}Time\\ (UT)\end{tabular} & Target & DIT & NDIT & Seeing ('') & $\tau_0$ (ms)\\
\hline
\hline
1 & 2018-01-27 & 05:30:21 & R Car & 3 & 100 & 0.76 & 5.4\\
2 &  & 05:42:30 & R Car & 3 & 100 & 0.63 & 5.5\\
3 &  & 06:25:34 & R Car & 3 & 100 & 0.81 & 5.2\\
4 &  & 06:37:45 & R Car & 3 & 100 & 0.89 & 6.0\\
5 &  & 06:51:46 & HD80404 & 30 & 10 & 0.76 & 6.1\\
6 &  & 07:03:09 & HD80404 & 30 & 10 & 0.99 & 5.8\\
7 &  & 07:29:58 & R Car & 3 & 100 & 0.73 & 6.3\\
8 &  & 07:42:04 & R Car & 3 & 100 & 0.67 & 5.0\\
9 &  & 07:56:18 & HD80404 & 30 & 10 & 0.85 & 5.0\\
10 &  & 08:07:34 & HD80404 & 30 & 10 & 0.90 & 4.3\\
11 &  & 08:22:08 & R Car & 3 & 100 & 0.50 & 4.5\\
12 &  & 08:34:15 & R Car & 3 & 100 & 0.58 & 4.4\\
\hline
13 & 2018-02-22 & 05:48:54 & R Car & 3 & 100 & 0.47 & 12.3\\
14 &  & 06:01:00 & R Car & 3 & 100 & 0.63 & 9.8\\
15 &  & 06:17:41 & HD80404 & 30 & 10 & 0.52 & 8.6\\
16 &  & 06:29:02 & HD80404 & 30 & 10 & 0.65 & 11.8\\
 \hline
17 & 2018-02-25 & 05:35:44 & R Car & 3 & 100 & 0.33 & 11.6\\
18 &  & 05:47:51 & R Car & 3 & 100 & 0.49 & 10.8\\
19 &  & 06:02:42 & HD80404 & 30 & 10 & 0.45 & 7.7\\
20 &  & 06:14:04 & HD80404 & 30 & 10 & 0.55 & 6.2\\
 \hline
21 & 2018-03-12 & 00:51:56 & R Car & 3 & 100 & 0.79 & 4.3\\
22 &  & 01:04:06 & R Car & 3 & 100 & 0.74 & 5.4\\
23 &  & 01:10:39 & R Car & 3 & 100 & 0.76 & 5.4\\
24 &  & 01:25:50 & HD80404 & 30 & 10 & 0.74 & 6.0\\
25 &  & 01:37:13 & HD80404 & 30 & 10 & 0.78 & 6.4 \\
\hline
\end{tabular}
\label{table:log_obs}
\end{table*}

\begin{figure*}
   \centering
   \subfigure[]{\includegraphics[width=17cm]{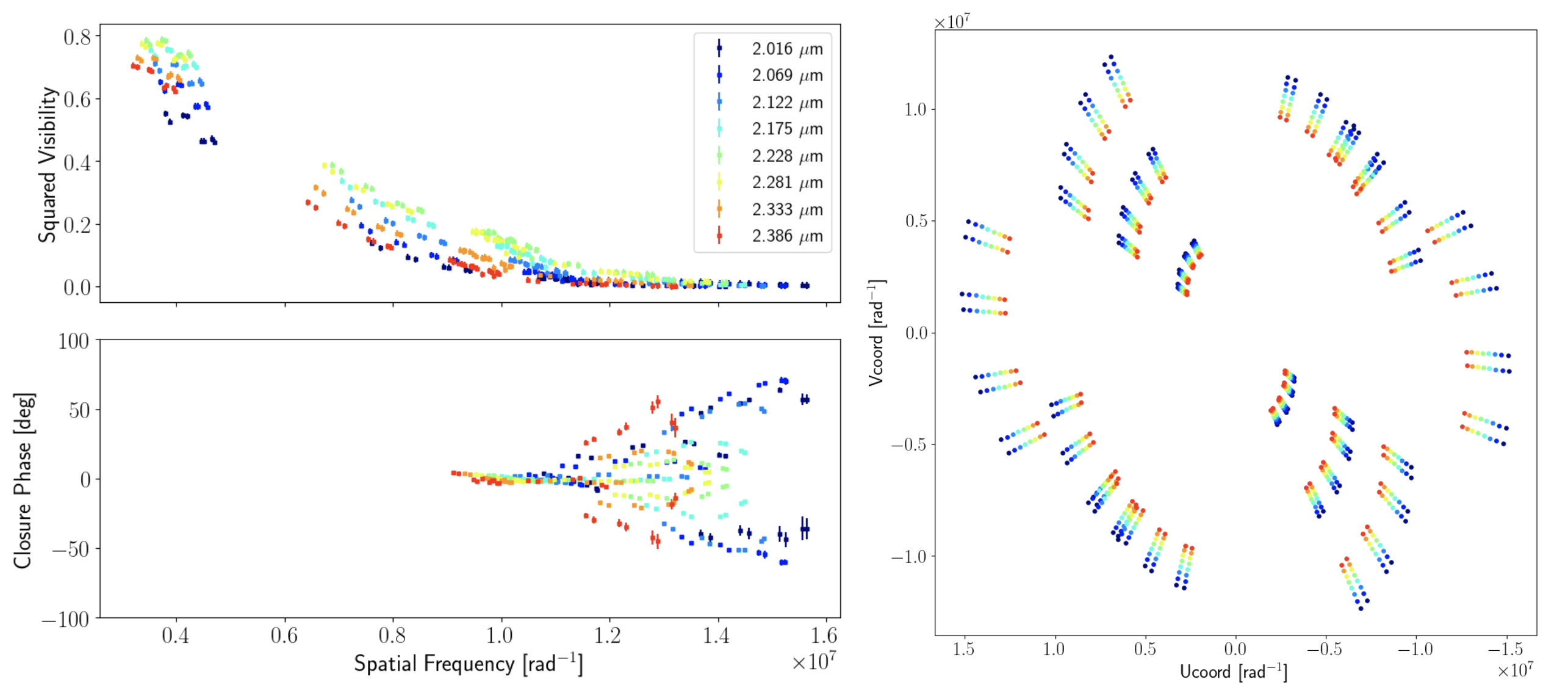}\label{subfig:vis2t3sf_uv_jan}}
   \subfigure[]{\includegraphics[width=17cm]{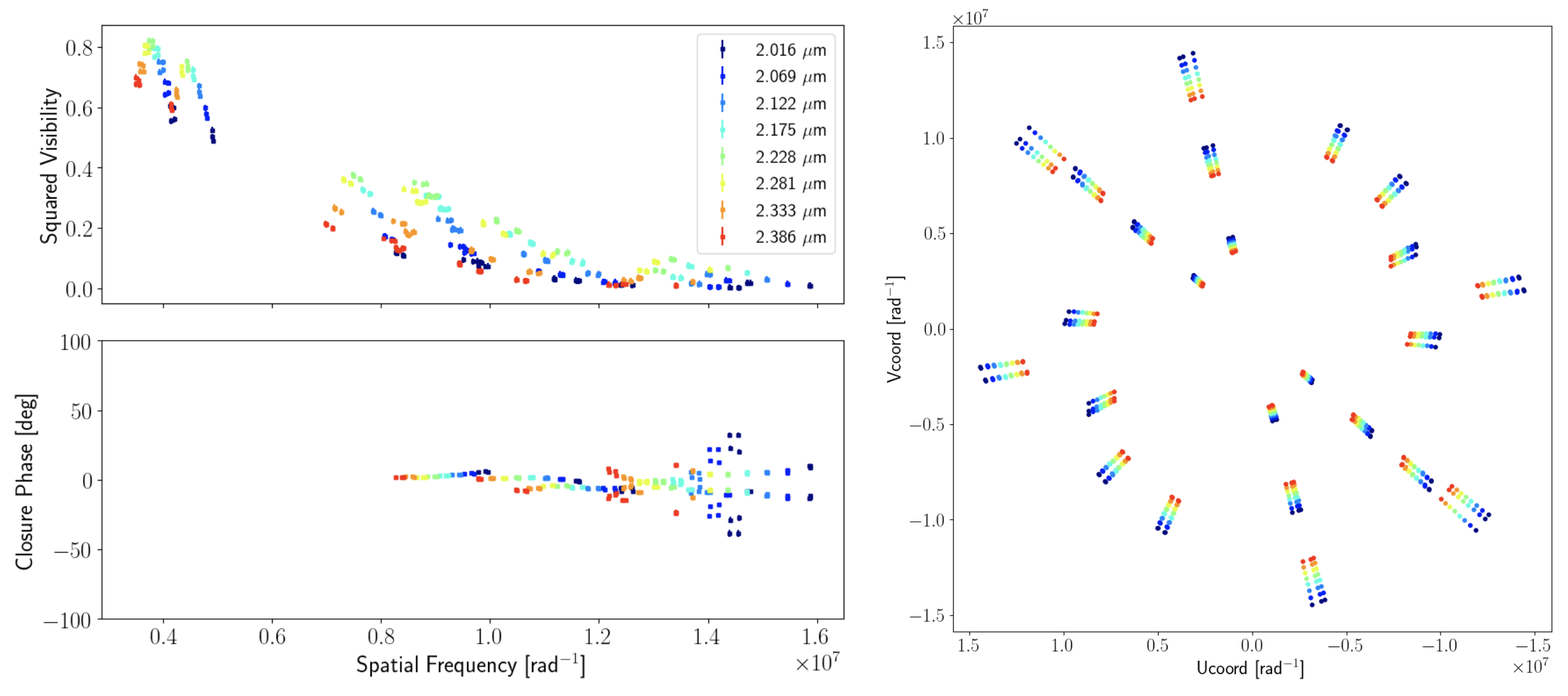}\label{subfig:vis2t3sf_uv_febmar}}
   
      \caption{\textbf{Left:} V$^2$ and CPs plotted versus spatial frequencies for the data sets obtained with GRAVITY. \textbf{Right:} The corresponding u-v coverage. Panel (a) corresponds to the January epoch while panel (b) corresponds to the February one. The colors on the plots show different wavelengths (see the labels on the plot).
              }
         \label{vis2t3sf_uv_jan}
   \end{figure*}
\section{Best-fit geometrical models}
In this section, we show the best-fit images to the observed V$^2$ and CPs across the pseudo-continuum, the 1st and 2nd CO band heads for our January and February epochs (see Sec. \ref{subsec:reconstruct_imgs}). In all figures, the red dots in the main panels correspond to the synthetic V$^2$ and CPs extracted from the raw reconstructed images, while the data are shown with gray dots in the main panels. The lower panels show the corresponding residuals (in terms of the number of standard deviations) coming from the comparison between the data and the best-fit reconstructed images.

\begin{figure*}
    \centering
    \subfigure[]{\includegraphics[width=9.1cm]{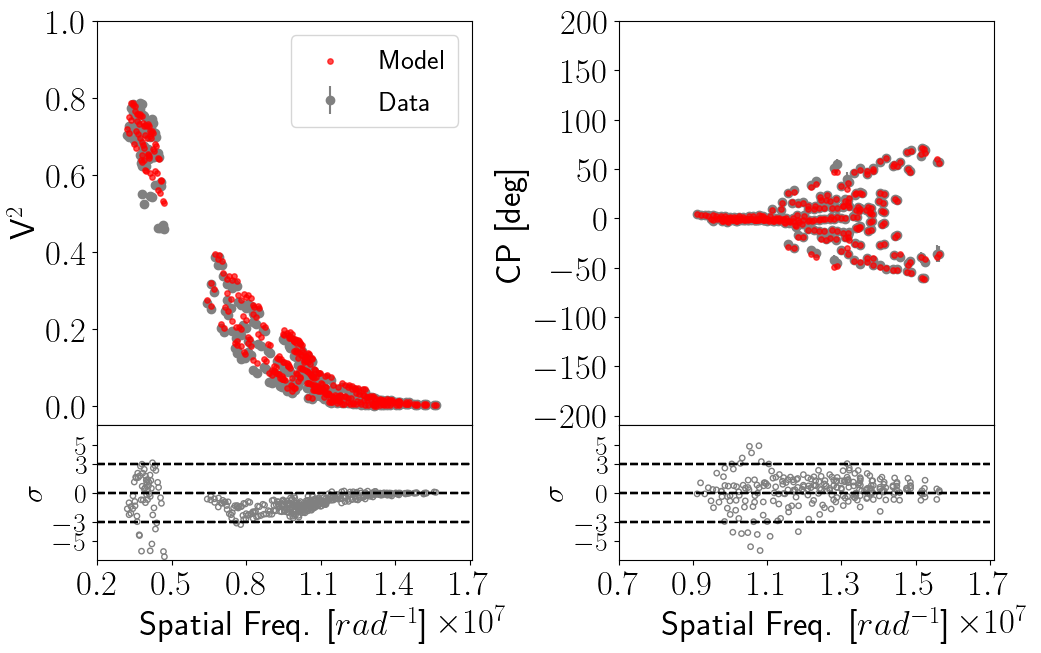}}
    \subfigure[]{\includegraphics[width=9.1cm]{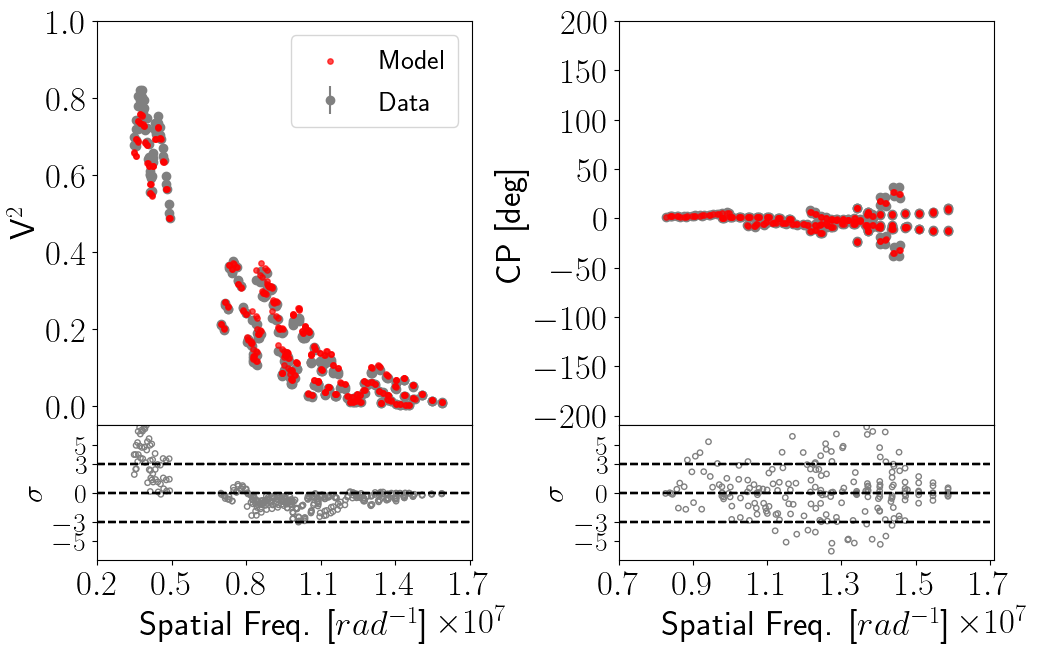}}
    \caption{Best-fit to the observed V$^2$ and CPs from the best reconstructed images across the pseudo-continuum for (a) the January and (b) February epochs. The red dots in the main panel correspond to the synthetic V$^2$ and CPs extracted from the raw reconstructed images. The data are shown with gray dots in the main panels. The lower panels show the residuals (in terms of the number of standard deviations) coming from the comparison between the data and the best-fit reconstructed images.}
    \label{fig:vis2_obs_jan_cont}
\end{figure*}

\begin{figure*}
    \centering
    \subfigure[]{\includegraphics[width=9.1cm]{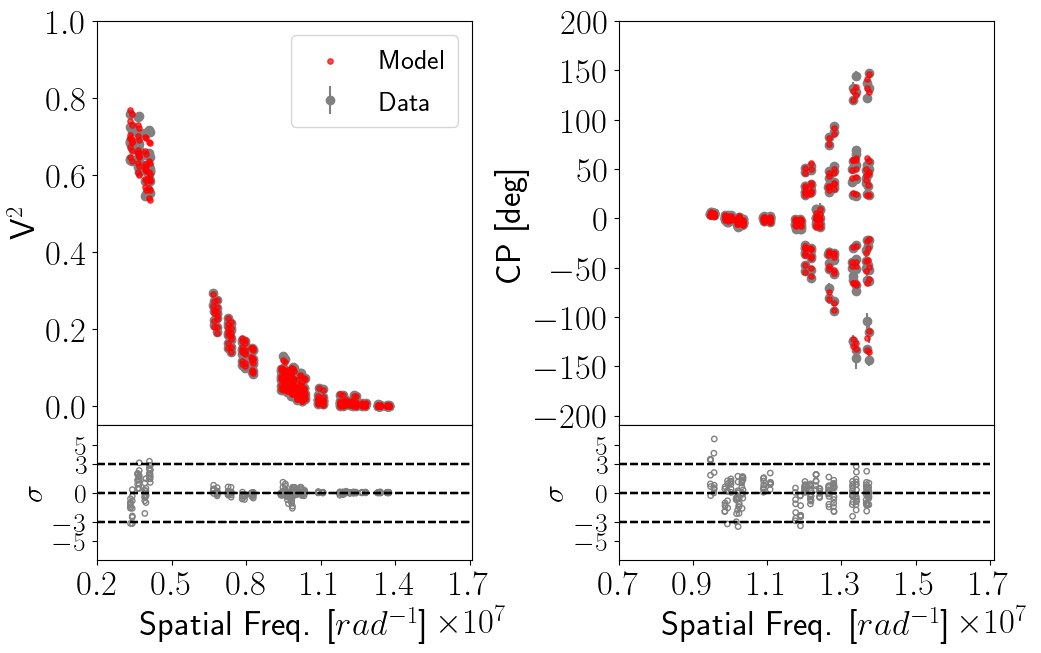}}
    \subfigure[]{\includegraphics[width=9.1cm]{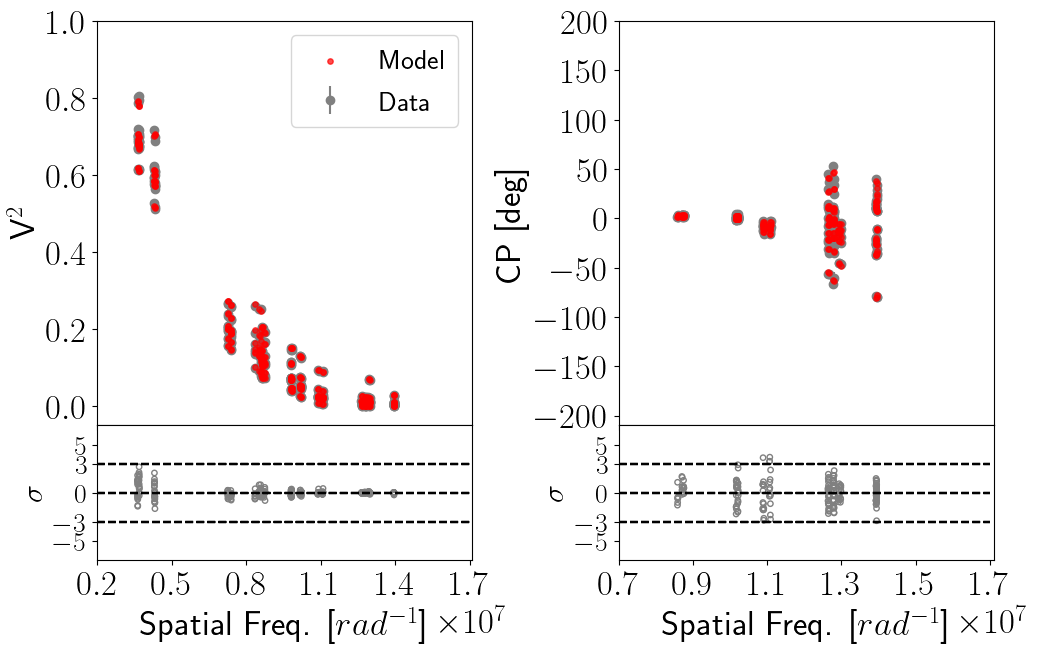}}
    \caption{Fit to the observed V$^2$ and CPs from the best reconstructed images across the 1st CO band head for a) the January  and b) the February epochs. The panels are as described in Figure \ref{fig:vis2_obs_jan_cont}.}
    \label{fig:t3_obs_jan_cont}
\end{figure*}

\begin{figure*}
    \centering
    \subfigure[]{\includegraphics[width=9.1cm]{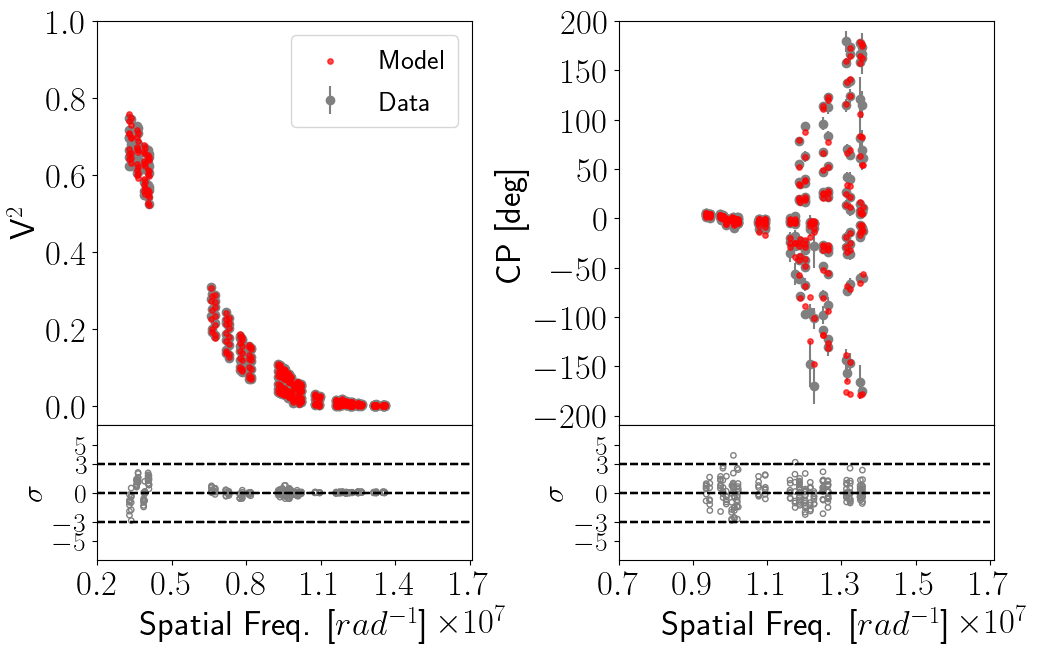}}
    \subfigure[]{\includegraphics[width=9.1cm]{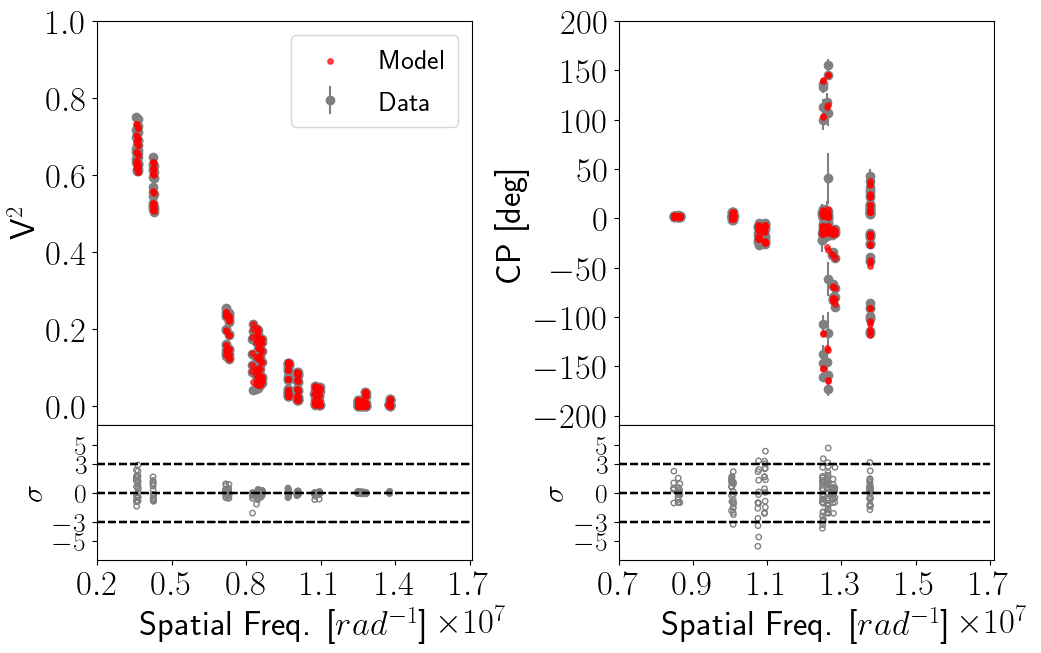}}
    \caption{Fit to the observed V$^2$ and CPs from the best reconstructed images across the 2nd CO band head for a) the January and b) the February epochs. The panels are as described in Figure \ref{fig:vis2_obs_jan_cont}.}
    \label{fig:vis2_obs_febm_cont}
\end{figure*}

\end{appendix}

\end{document}